\documentclass[sigconf]{acmart}

\AtBeginDocument{%
  }

\copyrightyear{2024}
\acmYear{2024}
\setcopyright{acmlicensed}\acmConference[CCS '24]{Proceedings of the 2024 ACM SIGSAC Conference on Computer and Communications Security}{October 14--18, 2024}{Salt Lake City, UT, USA}
\acmBooktitle{Proceedings of the 2024 ACM SIGSAC Conference on Computer and Communications Security (CCS '24), October 14--18, 2024, Salt Lake City, UT, USA}
\acmDOI{10.1145/3658644.3670323}
\acmISBN{979-8-4007-0636-3/24/10}

\usepackage{balance}

\usepackage{color}

\usepackage{tabularray}
\UseTblrLibrary{booktabs}

\usepackage{wasysym}

\usepackage{listings}
\usepackage{listings}
\usepackage{xcolor}

\definecolor{verylightgray}{rgb}{.97,.97,.97}
\definecolor{lightgray}{rgb}{.85,.85,.85}

\lstdefinelanguage{ABNF}{
    keywords ={},
    morecomment=[l]{;}, 
    morecomment=[s]{/*}{*/},
	commentstyle=\color{gray}\ttfamily,
}

\usepackage{xcolor}

\colorlet{punct}{red!60!black}
\definecolor{background}{HTML}{EEEEEE}
\definecolor{delim}{RGB}{20,105,176}

\lstdefinelanguage{json}{
    basicstyle=\footnotesize\ttfamily,
    numbers=left,
    numberstyle=\footnotesize,
    stepnumber=1,
    numbersep=5pt,
    showstringspaces=false,
    breaklines=true,
    frame=lines,
    backgroundcolor=\color{background},
    literate=
     *{:}{{{\color{punct}{:}}}}{1}
      {,}{{{\color{punct}{,}}}}{1}
      {\{}{{{\color{delim}{\{}}}}{1}
      {\}}{{{\color{delim}{\}}}}}{1}
      {[}{{{\color{delim}{[}}}}{1}
      {]}{{{\color{delim}{]}}}}{1},
}
\usepackage{subfigure}
\usepackage{pifont}
\usepackage{enumitem} 
\usepackage{xcolor}
\usepackage{hyperref}

 \hypersetup{
    colorlinks = true,
    linkcolor = blue,
    anchorcolor = blue,
    citecolor = red,
    filecolor = blue,
    urlcolor = purple
}

\usepackage{xspace}
\newcommand{\attack}{\textsf{\small Blind Message Attack}\xspace}
\newcommand{\attacks}{\textsf{\small Blind Message Attacks}\xspace}
\newcommand{\rattack}{\textsf{\small Replay Attack}\xspace}
\newcommand{\mattack}{\textsf{\small Blind Multi-Message Attack}\xspace}
\newcommand{\rattacks}{\textsf{\small Replay Attacks}\xspace}
\newcommand{\mattacks}{\textsf{\small Blind Multi-Message Attacks}\xspace}

\newcommand{\detector}{\textsc{Web3AuthChecker}\xspace}
\newcommand{\guard}{\textsc{Web3AuthGuard}\xspace}
\newcommand{\checker}{\textit{Checker}\xspace}
\newcommand{\fr}{\textit{FlexRequest}\xspace}

\newcommand{\fdomain}{\texttt{domain}\xspace}
\newcommand{\fname}{\texttt{name}\xspace}
\newcommand{\fnonce}{\texttt{nonce}\xspace}

\renewcommand{\paragraph}[1]{\vspace{3pt}\noindent{\bf{#1}.}}
\newcommand{\sparagraph}[1]{{\it{#1}.}}

\definecolor{red}{RGB}{215, 51, 51}
\definecolor{green}{RGB}{0, 171, 79}

\newcounter{notesec}[section]

\begin{document}

\title[Stealing Trust: Unraveling Blind Message Attacks in Web3 Authentication]{Stealing Trust: Unraveling Blind Message Attacks \\ in Web3 Authentication}

\author{Kailun Yan}
\authornote{Part of Kailun Yan's work was done when visiting George Mason University.}
\affiliation{%
  \institution{School of Cyber Science and Technology, Shandong University}
  \city{Qingdao}
  \country{China}
}
\orcid{0000-0003-0216-9793}
\email{kailun@mail.sdu.edu.cn}

\author{Xiaokuan Zhang}
\authornote{Corresponding authors.}
\affiliation{%
  \institution{Department of Computer Science, George Mason University}
  \city{Fairfax, VA}
  \country{United States}
}
\orcid{0000-0002-4646-7146}
\email{xiaokuan@gmu.edu}

\author{Wenrui Diao}
\authornotemark[2]
\affiliation{%
   \institution{School of Cyber Science and Technology, Shandong University}
  \city{Qingdao}
  \country{China}
}
\orcid{0000-0003-0916-8806}
\email{diaowenrui@link.cuhk.edu.hk}


\renewcommand{\shortauthors}{Kailun Yan, Xiaokuan Zhang, and Wenrui Diao}

\begin{abstract}
As the field of Web3 continues its rapid expansion, the security of Web3 authentication, often the gateway to various Web3 applications, becomes increasingly crucial. 
Despite its widespread use as a login method by numerous Web3 applications, the security risks of Web3 authentication have not received much attention. 
This paper investigates the vulnerabilities in the Web3 authentication process and proposes a new type of attack, dubbed {\it blind message attacks}.
In blind message attacks, attackers trick users into blindly signing messages from target applications by exploiting users' inability to verify the source of messages, thereby achieving unauthorized access to the target application.
We have developed Web3AuthChecker, a dynamic detection tool that interacts with Web3 authentication-related APIs to identify vulnerabilities.
Our evaluation of real-world Web3 applications shows that a staggering 75.8\% (22/29) of Web3 authentication deployments are at risk of blind message attacks.
In response to this alarming situation, we implemented Web3AuthGuard on the open-source wallet MetaMask to alert users of potential attacks.
Our evaluation results show that Web3AuthGuard can successfully raise alerts in 80\% of the tested Web3 authentications.
We have responsibly reported our findings to vulnerable websites and have been assigned two CVE IDs.
\end{abstract}

\ccsdesc[500]{Security and privacy~Authentication}

\keywords{Web3, Authentication, Blockchain, Security}

\maketitle

\section{Introduction}
Blockchain technology has been evolving rapidly since its first introduction by Satoshi Nakamoto in 2008~\cite{bitcoin}.
Web3 applications, taking advantage of the decentralized nature of blockchain technology, have garnered significant attention from both investors and users~\cite{DBLP:journals/tdsc/LiuXSGWXWLH22}.
Decentralized Finance, or DeFi, is a prominent example of the Web3 application, and its Total Value Locked (TVL) was around 52 billion USD as of December 2023~\cite{defiTVL2023}.
The total value of the Non-Fungible Token (NFT) marketplaces has exceeded two billion USD~\cite{nftcap}. 
In January 2024, the total value of assets in the Web3 game \textit{Midas Miner} was over one billion USD~\cite{MidasMiner}.
The considerable financial activities underscore the enormous potential of the Web3 ecosystem.

Web3 applications need to perform user authentication to grant the correct permissions for end users to access off-chain resources.
Web3 authentication is a challenge-response protocol where the user is identified by a public key (wallet address):
the Web3 application sends a specific message to the crypto wallet, signed by the user, and sent back to the application. 
The application verifies the signature and authenticates the user if it is valid~\cite{DBLP:journals/corr/abs-2304-06111}. This decentralized approach differs markedly from the traditional authentication protocols of Web2.
The most common way to perform Web3 authentication is through MetaMask~\cite{metamask}, 
a famous Web3 wallet for users to manage their digital assets and perform DeFi transactions.
As per statistics, MetaMask had 30 million monthly active users~\cite{metamask-stat}.
Since the Web3 application is fundamentally a website, in this paper, we use the terms \textit{application} and \textit{website} interchangeably.

Due to the popularity of Web3 and the large amount of money involved, the security of Web3 is paramount.
Existing research mainly focuses on the security risks associated with Web3 applications.
These include vulnerabilities in smart contracts
~\cite{DBLP:conf/ccs/SmolkaGWDDKP23, DBLP:conf/ndss/SendnerCFPKSDSK23, DBLP:conf/uss/HeZ00L0YJ21, DBLP:conf/sp/Bose0C0KV22, DBLP:conf/uss/ZhangZZL20, DBLP:conf/ccs/0002ZLLWCXZ19,DBLP:conf/sp/SoLPLO20, DBLP:conf/ndss/RodlerLKD19,  DBLP:conf/uss/SoHO21, DBLP:conf/uss/FrankAH20}, 
malicious DeFi attacks
~\cite{DBLP:conf/ccs/BabelJ0KKJ23, DBLP:conf/sp/ZhouXECWWQWSG23, DBLP:conf/uss/McLaughlinKV23, DBLP:conf/sp/BabelDKJ23, DBLP:conf/www/VictorW21, DBLP:conf/sp/ZhouQCLG21, DBLP:conf/fc/QinZLG21, DBLP:conf/sp/DaianGKLZBBJ20},
and the security of NFT marketplaces~\cite{DBLP:conf/ccs/0002BRKV22, DBLP:conf/www/WhiteMP22}.
The security and anonymity of crypto wallets~\cite{DBLP:conf/acns/SentanaIK23, DBLP:conf/www/YanZLDG23, DBLP:conf/securecomm/UddinMY21}, as well as cryptocurrency scams~\cite{DBLP:conf/ndss/LiYN23, DBLP:conf/ccs/HeCCHHWCWZ23} have also raised concerns.
However, to the best of our knowledge, 
almost no work has paid attention to the security of Web3 authentication.

To bridge the gap, this paper takes the {\it first} step toward understanding the security risks of Web3 authentication. 
We find that 
during the Web3 authentication process, there is no proper way for the user to identify the origin of the message to be signed.
Also, users use crypto wallets to manage key pairs, and one key pair is often used across multiple applications~\cite{DBLP:journals/toit/ChenLZCLLLZ20}.
As a result,
a malicious application A can present a valid signing message from application B for the user to sign (assuming the user uses the same key pair on both applications, which is very common~\cite{DBLP:journals/toit/ChenLZCLLLZ20}).
The user {\it blindly} signs the message, unwittingly granting application A the authorization to access application B.
We call such attacks \textit{\textbf{\attacks}}.
We have systematically analyzed the attack surfaces during the Web3 authentication process,
which leads to the identification of
several categories of potential vulnerabilities 
in the message design and server verification.


\paragraph{\detector}
To detect \attacks,
we have designed and implemented \detector, a dynamic detection tool that examines the security of the Web3 authentication process.
\detector bypasses the web page (front-end) and directly tests the API (back-end). 
It requests Web3 authentication-related APIs and checks the responses to find vulnerabilities.
The tool consists of two modules: 
1) \checker, which defines a set of general attack payloads and rules for testing APIs to find potential vulnerabilities,
and 2) \fr, an HTTP library designed to streamline testing identical APIs on different websites.
%
Specifically, to test a website, we load the authentication-related APIs of the website to \checker. \checker then instructs \fr to make a series of requests to these APIs.
Finally, \checker examines the data returned by \fr to identify potential vulnerabilities.


To evaluate the impact of \attacks, we collected 29 real-world cases of Web3 authentication, including 25 cases of using Web3 authentication for user login and 4 cases for profile update.
The 29 cases are from 27 websites, with some encompassing both types.
The websites we studied, which span marketplaces, games, and services, were sourced from DappRadar and Google searches. 
In January 2024 alone, these sites reported a total transaction volume exceeding 592 million US dollars and over 1.29 million unique active wallets (UAW), underscoring the widespread nature of these vulnerabilities.
By testing the APIs of 29 cases, \detector reports that 75.8\% (22/29) of Web3 authentications belonging to 20 websites are vulnerable to \attacks.
We manually checked all 29 cases and confirmed that the results were accurate.
Building upon the foundation of the \attack, we have developed more advanced attacks: the \rattack renders the session mechanism ineffective, while the \mattack allows attackers to acquire user identities across multiple websites in a single attack. In our study of 29 cases, we identified 11 instances of the \rattacks and 7 of the \mattacks.


\paragraph{\guard}
To address these vulnerabilities, websites and crypto wallets must implement a complete Web3 authentication update, but this is difficult to implement immediately.
Therefore, we proposed a user-side mitigation solution, \guard, to help users immediately mitigate \attacks in crypto wallets.
When a user logs into a website B, the wallet equipped with \guard will extract a message template from B's message. Later, when the user attempts to log into a new website A, the wallet will perform a regex match between the message from A and the message template from B. If a match is successful, the wallet will alert the user about the potential \attack on that website.
We implement \guard in MetaMask's open-source code.
We tested 25 user logins, and \guard successfully addressed 20 of them. 
\guard is unable to handle the remaining five cases because their vulnerabilities allow an attacker to modify the message body at will.




\paragraph{Responsible Disclosure}
Following the responsible disclosure policy, we reported the vulnerabilities we discovered to the corresponding websites. One of them (\textit{LearnBlockchain}) has acknowledged and fixed the vulnerability.
Additionally, since some vendors did not respond to our security reports, we submitted our findings directly to the CVE program, and two CVE IDs (CVE-2023-50053 and CVE-2023-50059) have been assigned.

\paragraph{Demos} The PoC demos of \attacks and mitigation are available at \url{https://sites.google.com/view/web3auth}.

\paragraph{Contributions} Our contributions are as follows:
\begin{itemize}[leftmargin = 12pt]
    \item \textbf{New Vulnerabilities and Attacks.} We conducted an in-depth investigation of the vulnerabilities in Web3 authentication and identified a new type of attack, called \attacks, and two advanced attacks -- \rattacks and \mattacks~(Section~\ref{sec:attacks}).
    \item \textbf{New Detection Tool (\detector).} We implemented a dynamic detection tool, \detector, to detect such vulnerabilities.
    It sends requests to Web3 backends and checks the responses to identify vulnerabilities (Section~\ref{sec:design}).
    \item \textbf{Empirical Analysis.}  We performed detection on 29 Web3 authentication cases using \detector, and found that 75.8\% of them are at risk of \attacks.
    We also conducted two case studies~(Section~\ref{sec:measurement}).
    \item \textbf{User-side Mitigation (\guard).} We designed and implemented a user-side mitigation solution, \guard, 
    which helps alert users about a potential \attack when the user attempts to log into a new website~(Section~\ref{sec:mitigation}). 
    \item \textbf{Open-source Release.} To benefit the Web3 community, we will open-source our code on GitHub, including the detection tool \detector\footnote{\url{https://github.com/d0scoo1/Web3AuthChecker}} and a MetaMask-based crypto wallet with built-in \guard\footnote{\url{https://github.com/d0scoo1/Web3AuthGuard}}.
\end{itemize}

  
\begin{figure*}[t]
    \centering
	\includegraphics[width=0.95\linewidth]{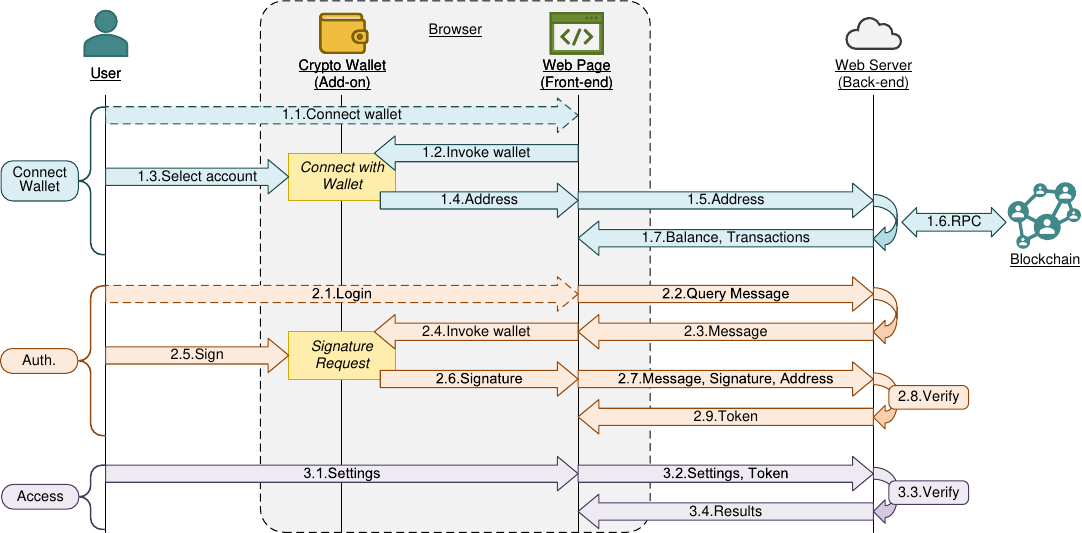}
	\caption{Web3 Authentication Process.}
	\label{fig:web3_auth}
\end{figure*}

\section{Background}
\label{sec:background}
This section introduces the background of Web3, Web3 authentication, and crypto wallets, then explores the potential impacts of unauthorized access in Web3 applications.

\subsection{Web3}
Web3, also known as Web 3.0, signifies the upcoming evolution of the Internet. It will move away from centralized servers and data centers owned by a few large corporations and instead use a distributed network, blockchain, to host the data ~\cite{DBLP:journals/tdsc/LiuXSGWXWLH22}.
Blockchains are decentralized ledgers that record transactions across many nodes (miners). 
This ensures that the data cannot be altered without the consensus of the network.

\paragraph{Web3 Application}
Applications built on the blockchain network are called Web3 applications or decentralized applications (DApps). 
Web3 applications cover a broad scope, including NFT marketplaces (NFTM), which allow creators and buyers to trade digital assets efficiently and securely; 
decentralized exchanges (DEX), which enable trading of digital currencies without a central authority; 
and decentralized finance (DeFi) platforms provide financial services such as lending and borrowing. 
Each Web3 application typically consists of a smart contract and a website. 
Smart contracts automate operations (transactions) and store critical data on the blockchain, enhancing transparency and security. 
Websites serve as the user interface and manage off-chain data. 
While some Web3 applications are embedded within mobile apps like crypto wallets, they fundamentally function as websites. 
This paper focuses on websites of Web3 applications and uses the terms \textit{application} and \textit{website} interchangeably.

\paragraph{Web3 Authentication}
Web3 authentication is a decentralized off-chain authentication technology based on cryptography. 
Web3 authentication offers benefits such as anonymity, security, and decentralization and is widely used in Web3 applications~\cite{Web3Auth}. In these applications, users can employ their public keys as identifiers to manage their off-chain data.
Specifically, it uses asymmetric encryption, where the client (crypto wallet) signs a specific message with their private key, and the server (Web3 application) verifies the message and its signature to authenticate the user’s identity (public key)~\cite{DBLP:journals/ijisec/JohnsonMV01}. 
Compared to Web2 authentication technologies, Web3 employs elliptic curve algorithms, which provide enhanced security. Moreover, in the Web3 ecosystem, public keys serve as identifiers, offering better anonymity. 
In contrast, Web2 authentication methods~\cite{DBLP:conf/raid/PhilippaertsPJ22, DBLP:conf/uss/MunyendoMNGKUA22, DBLP:conf/uss/GollaHLPR21, DBLP:conf/ccs/JannettMMS22} often require users to provide personally identifiable information (PII), such as phone numbers and email addresses. These are stored on centralized servers, exposing users to risks like privacy breaches.
Furthermore, Web3 authentication allows users to maintain a unified identity (public key) across multiple applications and services. 
Users no longer need to manage accounts and passwords for each website, which simplifies the user experience and reduces security risks.

\paragraph{Crypto Wallet}
Crypto wallets are a crucial component of the Web3 ecosystem, allowing users to securely store and manage their digital assets (key pairs) using hardware or software~\cite{DBLP:conf/securecomm/UddinMY21}. 
Most Web3 applications (websites) are compatible with popular crypto wallets and interact with them through APIs such as {\it Web3.js}~\cite{web3js} or {\it ethers.js}~\cite{ethers}.
When a user accesses a website, the website examines the browser for the presence of a Web3 provider, which is the API interface for crypto wallets. If the Web3 provider is detected, the website establishes a connection to it and prompts the user to connect their wallet. 
After user approval, the wallet shares the account addresses (public keys) with the website. This facilitates subsequent interactions, including transactions, smart contract operations, and Web3 authentication.

\subsection{Impacts of Unauthorized Access in Web3 }
\label{sec:attack impact}

\paragraph{Asset Loss}
Web3 authentication only grants users access to off-chain data, and on-chain transactions usually require additional signatures so that Web3 authentication will not impact users' on-chain assets.
However, attackers can still profit from unauthorized access to user accounts.
Here, we highlight two significant asset losses: \textit{Unlocked Content} and \textit{Unfair Trading}.

\sparagraph{Unlocked Content}
In NFT marketplaces, some items are valued from \textit{unlocked content}. If attackers gain unauthorized access to these accounts, they can directly access the \textit{locked content} without any cost. Moreover, most Web3 websites do not have abnormal login detection, meaning that the owners may never be notified when their locked content is leaked or stolen.

\sparagraph{Unfair Trading} 
An NFT's characteristics directly influence its value, with rarer traits often commanding higher prices~\cite{DBLP:conf/ccs/0002BRKV22}. For instance, in a popular game, CryptoKitties, the cost of each cat is determined by 13 distinct properties. 
While an attacker generally cannot manipulate the price of an item without signature authorization, they could alter the properties without restrictions in collections that support \textit{lazy minting}. 
This could allow an attacker to change the properties of a low-value item to a rare one and then purchase it, resulting in an \textit{unfair trade}. 
Even if the owner detects the attack afterward, they cannot reverse the transaction.
Lazy minting allows creators to produce NFTs without upfront gas costs, significantly reducing the barriers to selling NFTs.

\paragraph{Compromised Anonymity and Reputation}
The key pair in the crypto wallet is randomly generated, so the user's public key (address) has good anonymity, but an attacker who gains access to a user's account can link the user's address to personal information such as email, effectively breaking the anonymity.
Moreover, unauthorized access can damage users' reputations, as attackers may use the user's account for illegal activities.
NFT marketplaces allow users to link their social media accounts, such as Twitter or Facebook.
These social media accounts represent important channels for artists to strengthen their connections with users and promote their artworks~\cite{DBLP:conf/www/WhiteMP22}, which attackers can exploit to encourage illegal activities, causing reputational damage.
%

\section{Web3 Authentication}
\label{sec:web3 auth}
This section briefly introduces the Web3 authentication process, then specifically focuses on \textit{message design} and \textit{server verification} in authentication.

\subsection{Overview}
Web3 authentication is primarily utilized in two scenarios: \textit{user login} and \textit{profile update}. 
In the first scenario, the client needs to sign a specific message, which is then verified by the server. 
Once authenticated, the server will issue a token to the client to sustain a session.
In the second scenario, the user signs the updated profile, and the server checks this signature before updating the user's profile.
These two scenarios are independent; some websites may include only one of them, while others incorporate both. 
The user login is the most common scenario of Web3 authentication, and therefore, this paper mainly focuses on this scenario.
Figure~\ref{fig:web3_auth} illustrates the process of a user logging into a website by Web3 authentication. The process is described below.

\paragraph{Connect Wallet}
Connecting a wallet with the website is the first step for a user to interact with a Web3 website.
The user clicks the \textit{Connect Wallet} button on the web page (front-end) \textsf{\small (1.1)}, then the wallet pops up a window asking the user to select the account they want to connect to the website \textsf{\small (1.2)}.
The user selects an account (address, i.e., public key) \textsf{\small (1.3)}, and the wallet returns the address to the web page \textsf{\small (1.4)}. The web page forwards this address to its server (back-end) \textsf{\small (1.5)}. The server, using RPC, fetches the balance and transactions of the address from the blockchain \textsf{\small (1.6)}, then returns this information to the web page \textsf{\small (1.7)}. The web page displays the user's balance and transactions on the page.
It is worth noting that while multiple accounts may be connected, the website will interact with the first account by default. For simplicity, we assume that only one account is selected in this paper.

\paragraph{Authenticate}
Upon the user clicking the \textit{login} button \textsf{\small (2.1)}, the web page requests a message from the server \textsf{\small (2.2)}. 
Then, the web page forwards the message to the wallet to prompt a signature request \textsf{\small (2.3-2.4)}. A window pops up from the wallet, asking the user to sign the message.
After the user approves the signature request \textsf{\small (2.5)}, the wallet returns the signature to the web page \textsf{\small (2.6)}. 
Then, the web page sends the signature, message, and address to the server for verification \textsf{\small (2.7-2.8)}. If authenticated successfully, the server returns a token to the web page \textsf{\small (2.9)}.

\paragraph{Access} 
The user (front-end) holds the token to access protected information \textsf{\small (3.1-3.4)}. To keep the session, the website may store the token in the request header or cookie. This paper only focuses on Web3 authentication, so the session details are omitted.

\begin{lstlisting}[language=ABNF, caption={ABNF Message Format}, label={lst:abnf}]
message = 1*field
    ; Message consists of multiple fields.

field = statement / domain / name / nonce / ext

statement = *( reserved / unreserved / " " )
    ; See RFC 3986 for the definition of "reserved" and "unreserved".

domain = authority
    ; See RFC 3986 for the definition of "authority".

name = *( ALPHA / DIGIT / "-" )
    ; Website Name.

nonce = timestamp / date-time / rnd
    ; See RFC 3339 for the definition of "date-time".

timestamp = 10( DIGIT ) / 13( DIGIT )

rnd = *( ALPHA / DIGIT )
    ; Random Number.

ext = address / version / chain-id / issued-at / expiration-time / not-before / request-id
    ; Extension Fields.
\end{lstlisting}

\begin{lstlisting}[language=ABNF, caption={A Message from Opensea.io (Good Design)}, label={lst:good message}]
Welcome to OpenSea! 

This request will not trigger a blockchain transaction or 
cost any gas fees. 
Click to sign in and accept the OpenSea Terms of Service:
https://opensea.io/tos

Your authentication status will reset after 24 Hours.

Wallet address: 0x36e7c6feb20a90b07f63863d09cc12c4c9f39064
Nonce: 66ffb8f1-5eb1-4477-9558-36a60eb1b51f
\end{lstlisting}

\begin{lstlisting}[language=ABNF, caption={A Message from Foundation.app (Bad Design)}, label={lst:bad message}]
Please sign this message to connect to Foundation.
\end{lstlisting}

\subsection{Message Design}
In Steps \textsf{\small 2.3-2.6}, Web3 authentication normally uses the protocol \textit{Personal Sign} (EIP-191)~\cite{ERC-191}. This protocol does not impose any requirement on the format or content of a message, so Web3 applications must implement their own messages for authentication.
We conducted an extensive survey and identified common fields in messages, represented in Augmented Backus–Naur Form (ABNF) as shown in Listing~\ref{lst:abnf}. The fields referenced from other RFCs~\cite{rfc} are annotated in the ABNF.
These fields can be categorized as variable fields, such as \fnonce and \texttt{address}, and static fields, like \fdomain and \texttt{statement}.
For convenience, the content composed of static fields in the message is referred to as the \textit{message body}.

In Web3 authentication, each field serves a distinct function, and the design of the message differs across various websites. Listing~\ref{lst:good message} and Listing~\ref{lst:bad message} provide two examples from well-known NFT marketplaces, \textit{opensea.io} and \textit{foundation.app}, respectively. The message from \textit{opensea.io} is well-designed, including fields such as \fdomain and \fnonce. In contrast, the message from \textit{foundation.app} contains only \fname and \texttt{statement}, thereby posing security vulnerabilities. Next, we will explain the meaning and function of each field in Listing~\ref{lst:abnf} in detail, using these two examples as references.

\paragraph{statement} The \texttt{statement} field provides a human-readable text that explains the message. For instance, the message in Listing~\ref{lst:bad message} contains the statement \textit{``Please sign\ldots''}, prompting users to sign.

\paragraph{domain} The \fdomain field is a critical element of the message, representing the website's domain name. It plays a pivotal role in message security, enabling users to verify the source of the message. For example, the message in Listing~\ref{lst:good message} includes the \fdomain field \textit{opensea.io}. If a malicious website prompts the user to sign this message, the user can quickly recognize that it comes from \textit{opensea.io} and refuse to sign it.

\paragraph{name} The \fname field, indicating the website name, offers identifiability but may be unreliable due to potential duplication.

\paragraph{nonce} The \fnonce field is vital for fortifying security against replay attacks. Nonces can be classified into two categories:

\begin{itemize}[leftmargin = 12pt]
\item \textit{Time-based nonces}, usually \textit{timestamp} or \textit{datetime}, are generated by the front-end and embedded into the message. The server then verifies if the nonce is within its acceptable time range.
\item \textit{Record-based nonces}, typically \textit{random numbers}, require the front-end to query the backend before signing. Record-based nonces can be further divided into \textit{one-time nonces} per address and \textit{temporary nonces}, reusable within a specific validity period.
\end{itemize}
The choice between these nonces depends on factors such as distribution and storage, and no absolute preference exists.


\paragraph{ext} The \texttt{ext} field serves as an extension and is not central to this work. For more details, please refer to EIP-4361~\cite{ERC-4361}.

\subsection{Server Verification}
Steps \textsf{\small 2.7-2.9} outline the verification process of Web3 authentication. The front-end submits the {\it message}, {\it signature}, and {\it address} to the server, which then scrutinizes these data to authenticate the user's authority to log into the website. The server's verification process typically involves the following three steps:

\paragraph{Static Field Verification}
The server first checks the validity of the message body (static fields). It disregards variable fields, such as \fnonce, and compares the remaining content of the message with its local message body. If a discrepancy exists between the two, the authentication process is terminated, returning a failure response.

\paragraph{Variable Field Verification} The server verifies the \fnonce in the message to avoid replay attacks. Depending on the nonce type, the server might need to query the existence of a nonce record (for record-based nonces) or verify whether the timestamp is within an acceptable time range relative to its internal clock (for time-based nonces).
Other variable fields on the server, such as \texttt{address}, will also be verified during this process.

\paragraph{Signature Verification} The server verifies the validity of the signature: it uses the user's address (public key) to decrypt the signature. Then, it compares the result with the hash of the message. If they match, the signature is deemed valid. Once verified, the server issues an authentication token to authorize user access.

\section{Blind Message Attack}
\label{sec:attacks}
This section presents \attacks, which exploit vulnerabilities in the Web3 authentication implementation described in Section~\ref{sec:web3 auth} to gain unauthorized access.
Specifically, when a user logs into a malicious website using Web3 authentication, it tricks the user into signing a message from another website, thereby gaining access to the user's account on that other website.
This section first defines the threat model, followed by a motivating example.
Then, we analyzed the potential vulnerabilities in the Web3 authentication process.
Finally, according to the impact of these vulnerabilities, we classify the risk levels of \attacks and discuss advanced attacks.
Attack demos are available at \url{https://sites.google.com/view/web3auth}.

\paragraph{Threat Model}
In our threat model, the attacker aims to steal users' identities on target websites, thereby gaining unauthorized access. This occurs when a user logs into the attacker's (malicious) website using Web3 authentication. The attacker exploits this opportunity to steal the user's identity (signature) of the target website. 
In our model, the attacker cannot exploit browser vulnerabilities or crack users' crypto wallets. 
Also, attackers cannot intercept and decrypt network packets, such as man-in-the-middle (MitM) attacks.
The attacker's capabilities are limited to exploiting vulnerabilities in the Web3 authentication process of the target website.

\begin{figure}[t]
    \centering
    \includegraphics[width=0.98\linewidth]{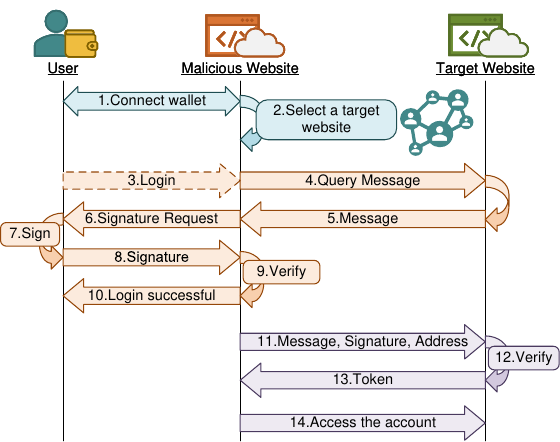}
    \caption{Blind Message Attack.}
    \label{fig:attack_flow}
\end{figure}

\subsection{Motivating Example}
Figure~\ref{fig:attack_flow} shows in detail how an attacker steals a user's identity on the target website when the user logs in using Web3 authentication on the malicious website.
The attack process is as follows:

\paragraph{(a) Select a Target Website} In Steps \textsf{\small 1-2}, a user visits a malicious website and connects to the wallet. 
The malicious website queries the transaction record of the address (public key) in the blockchain, identifies the websites that the user has logged into before by contract addresses, and then selects the most valuable target website based on information such as balance and trading volume.

\paragraph{(b) Login} In Steps \textsf{\small 3-6}, when the user logs in, the malicious website impersonates the user to obtain a message from the target website and prompts the user to sign it. Here, the malicious website can bypass the target website's front-end and obtain the message directly from the back-end through an HTTP request.
In Steps \textsf{\small 7-10}, the user signs the message and returns the signature, after which the malicious website then verifies the signature using the user's public key and prompts that the login is successful.

\paragraph{(c) Unauthorized Access} In Steps \textsf{\small 11-14}, the malicious website uses the signature, message, and address previously obtained to request a token from the target website and then holds the token to access the user's account.

\paragraph{Root Cause}
According to the above attack process, the root cause of the \attack lies in the user's action of blindly signing a message intended for a legitimate (target) website while on a malicious website during the Web3 authentication process.
The attacker then uses this leaked signature to bypass the target website's Web3 authentication.
Notably, the malicious website does not need to impersonate a specific target website — it can be any website that successfully entices users to log in. 
The success of a \attack hinges on \textit{the victim's unawareness that the message being signed is from another website}, and this vulnerability is dependent on the security flaws of the target website.

\subsection{Vulnerability Analysis}
\label{sec:vulnerability}
Vulnerabilities in the Web3 authentication process are crucial in facilitating \attacks. 
According to Section~\ref{sec:web3 auth}, these vulnerabilities stem from two primary sources:
1) flaws in message design resulting in the exclusion of essential fields (V1)
and 2) inadequate verification allows security measures to be circumvented (V2, V3).
Therefore, the vulnerabilities of Web3 authentication mainly include the following three aspects:

\paragraph{(V1) Lack of Essential Fields}
In the EIP-191 protocol, developers can customize the message in Web3 authentication.
Typically, \fdomain serves as an identifier of a message to indicate the issuer; if it is missing, users cannot determine the precise source of the message. Moreover, the potential duplication of \fname renders it unreliable as an identifier.
For instance, the \textit{foundation.app} message in Listing~\ref{lst:bad message} includes only the name \textit{Foundation}, enabling malicious sites (e.g., \textit{foundation.com}) to imitate legitimate ones. 
Furthermore, the lack of the \fnonce field makes it possible for a leaked signature to be reused for token refreshment, enabling replay attacks.

\paragraph{(V2) Unchecked Fields}
Failure of the server to check the fields in the message can cause security risks.
For instance, some servers fail to check static fields, such as the message body. This oversight allows malicious websites to alter fields like \fdomain or \fname, misleading users about the message's issuer. Additionally, neglecting to verify the \fnonce field opens the possibility of replay attacks.

\paragraph{(V3) Verification Flaws}
Verification flaws on the server allow specially crafted messages to evade checks. 
A common error is using regular expressions (regex) to match the message body with variable fields without confirming the message body's exact match with the issued message. 
Since regex operates on the principle of \textit{inclusion} rather than \textit{equality}, messages with extra content can also pass verification.
Moreover, risks may also arise when checking the time-based nonce, as an appropriate time range is not set.

\subsection{Security Risks}
This section discusses the impact of vulnerabilities on \attacks and discusses two advanced attacks, namely \rattacks and \mattacks.

\subsubsection{Risk Levels}
Based on the vulnerability analysis in Section~\ref{sec:vulnerability}, we categorize \attacks into four levels.
In critical to medium-risk scenarios, users are often unable to detect the attack or find it challenging to do so.
In low-risk scenarios, attentive users might identify the message as originating from another website, thereby preventing the attack.

\paragraph{Critical Risk} 
In critical-risk scenarios, the attacker can directly launch a \attack and steal the user's identity without even interacting with the victim (i.e., skipping \textsf{\small 3-7} in Figure~\ref{fig:attack_flow}).
This risk arises from the absence of verification for signatures, addresses, and messages. Without the server checking the signature or the address, an attacker only needs to set the address in the request as the user to impersonate them without requiring the user's signature. 
Suppose the server only verifies the signature and address but not the content of the message. In that case, the attacker does not need to obtain the user's specific signature on the website; he can impersonate the user using any of the user's messages and signatures.
\looseness=-1

\paragraph{High Risk}
In high-risk scenarios, the user would not be able to recognize that the malicious website's message came from another website, so once the user intends to log into the malicious website, the \attack will be successful.
This vulnerability arises from shortcomings in message design (V1) or server verification (V2). 
Specifically, if messages from the target website lack the \fdomain and \fname, users cannot identify the message's source in (malicious) websites. Furthermore, if the target website does not verify the message body, a malicious website can alter the \fdomain and \fname to its own, thereby misleading users.

\paragraph{Medium Risk}
In medium-risk scenarios, 
users may find that the malicious website's message comes from the target website and refuse to sign it, thereby avoiding the attack. 
Specifically, if the message from the target website contains \fname but lacks a \fdomain (V1), the malicious website has to register a similar domain name to mislead the user. 
If the message body check is sloppy (V3), such as only regex matching, attackers can hide the target website's message body within their own messages.
Although both situations may be noticed by experienced users, in fact, most users are not aware of it, as demonstrated by the second case in Section~\ref{sec:case study}.


\paragraph{Low Risk}
In low-risk scenarios,
the message contains a \fdomain, and there are no flaws in the server verification. An attack will only succeed if the user disregards the signed message. 
Due to the common lack of Web3 expertise among users and the similarity in messages across various websites, the attack might still succeed.

\subsubsection{Advanced Attacks}
Serious vulnerabilities in Web3 authentication also provide the basis for advanced attacks, specifically \rattack and \mattack. 
Through replay attacks, attackers can maintain prolonged unauthorized access. 
Through blind multi-message attacks, attackers can obtain a user's identity across various websites within a single \attack.

\paragraph{Replay Attack}
Generally, an authentication token has a specified expiration time, and once it expires, the attacker cannot access the user's account without securing a new signature. 
However, if the message lacks \fnonce (V1) or the verification process is flawed (V2, V3), an attacker can replay the signature repeatedly to refresh the token, thus maintaining unauthorized access.
It is important to note that while broader replay attacks could be carried out by intercepting signatures, such as man-in-the-middle attacks, such cases fall outside the threat model this paper considers. Our focus remains on assessing the impact of replay attacks within the framework of \attacks.

\paragraph{Blind Multi-Message Attack}
Attackers can exploit vulnerabilities across multiple websites to create a crafted message that simultaneously passes Web3 authentication on all these websites.
As a result, with a single signature from the user, the attacker can access the user's accounts on several websites. 
The crafted message is equivalent to multiple messages on websites, so we call the attack \mattack.
Specifically, for websites that do not verify the message body (V2) or only verify its presence rather than strict equality (V3), 
an attacker can bypass the server verification of these websites by constructing a message that contains the required fields.
We will describe this attack in detail in Section~\ref{sec:case study}.

\section{\detector}
\label{sec:design}
This section presents \detector, a tool for automatically detecting \attacks through dynamic analysis of target websites. We start with an overview of the architecture, followed by a detailed implementation of three checkers (message, nonce, and signature). Finally, we introduce \fr, an HTTP request tool designed to streamline the testing process.

\subsection{Overview}

As depicted in Figure~\ref{fig:attack_flow}, the interaction between a malicious website and a target website involves three requests:
\begin{itemize}[leftmargin = 12pt]
\item \texttt{QUERY}. In Step \textsf{\small 4}, the malicious website queries a message from the target website's server.
\item \texttt{AUTH}. In Step \textsf{\small 11}, the malicious website forwards the message, signature, and address to the target website's server to authenticate and obtain a token.
\item \texttt{ACCESS}. In Step \textsf{\small 14}, the malicious website holds the authentication token to access the user's account.
\end{itemize}

\begin{figure}[t]
    \centering
    \includegraphics[width=1\linewidth]{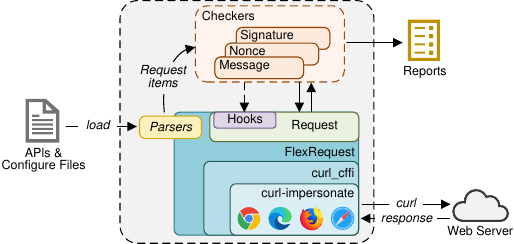}
    \caption{Architecture of Web3AuthChecker.}
    \label{fig:Web3AuthChecker}
\end{figure}

Our tool, \detector, interfaces with these requests to identify vulnerabilities. 
In simple terms, it injects different attack payloads into HTTP requests, then analyzes whether responses are as expected.
The architecture, depicted in Figure~\ref{fig:Web3AuthChecker}, is composed of \checker and \fr.
There are some predefined parsers in \fr to support converting different types of API and configuration files into a unified \textit{Request items} object.
The vulnerability detection process is as follows:
\begin{enumerate}[leftmargin=12pt]
\item First, parsers convert loaded APIs and configuration files into \textit{Request items} and send them to \checker.
\item \checker guides \fr to perform a series of API requests. For each request, \checker supplies attack payloads and API parameters to \fr. \fr utilizes its auto-replacement mechanism to insert the attack payloads into the proper slots within the API parameters, subsequently forwarding the request to the website under test.
\item \fr parses the website's response according to the predefined configuration and returns the parsed results back to \checker.
\item After completing a series of test requests, \checker scrutinizes the data returned by \fr to detect vulnerabilities.
\item Finally, \checker generates a test report for each website, highlighting any identified vulnerabilities.
\end{enumerate}

Through the process mentioned above,
\detector efficiently evaluates the security of Web3 authentication implementations on target websites, identifying potential risks. 
Subsequently, we will elaborate on the two components of \detector.


\subsection{Checker Implementation}
\label{sec:checker}
We developed three checkers to detect vulnerabilities in Section~\ref{sec:vulnerability}. Each checker focuses on one aspect of Web3 authentication -- message, nonce, and signature -- to identify potential vulnerabilities. The checkers assess the integrity of the message design by examining specific fields (V1) and employ various attack payloads to uncover vulnerabilities in verification (V2, V3).

\paragraph{Message Checker}
Message Checker begins by sending a \texttt{QUERY} request to retrieve a message from the website. 
Then, it matches the presence of domain names (\fdomain) and website names (\fname).
To detect vulnerabilities in verification, the checker modifies the message to a random string as the attack payload and signs it. If a subsequent \texttt{AUTH} request returns a token, it indicates that the server does not adequately check the message. To further verify, the checker 1) empties the message body and 2) appends random characters at the beginning of the message to assess any vulnerabilities in the server's message body verification.
If the message fails any of the above checks, it is considered vulnerable to \attacks, and the checker returns specific security risk levels based on the type of vulnerability.

\paragraph{Nonce Checker}
Nonce Checker sends \texttt{QUERY} requests from multiple addresses to obtain messages. It then compares these messages to identify variable fields, excluding non-nonce fields. 
If variable fields persist, it suggests the presence of \fnonce.
Subsequently, the checker executes the following sequence of requests:

\begin{enumerate}[leftmargin=12pt]
\item It sends a \texttt{QUERY} and \texttt{AUTH} request to the server.
\item It repeats the first \texttt{AUTH} request. 
\item It uses a new address to retrieve another message by a \texttt{QUERY} request. It then uses the old address to sign this new message and sends the third \texttt{AUTH}.
\item It generates a value similar to the nonce and replaces it before sending the fourth \texttt{AUTH}.
\item It removes the nonce before sending the fifth \texttt{AUTH}.
\end{enumerate}
If a token is still returned after these requests, it implies that the server has failed nonce verification.
The type of nonce can also be inferred from these requests, as detailed in Table~\ref{tab:check nonces}.

\begin{table}[t]
\caption{AUTH Request Status vs. Nonce Type}
\label{tab:check nonces}
\centering
\footnotesize
\SetTblrInner{colsep=6pt,rowsep=0pt,stretch = 1}
\begin{tblr}{
    width = 1\linewidth, 
    colspec ={X[2.5,l]X[1,c]X[1,c]X[1,c]X[1,c]X[1,c]},
    row{odd} = {white},
    row{even} = {gray!25},
    row{1} = {white, font=\bfseries},
    row{2} = {white, font=\bfseries},
    column{1} = {font=\bfseries},
    cell{1}{2} = {c=5}{c},
    cell{1}{1} = {r=2}{l},
}
\toprule
{Nonce Type}   &AUTH Request\\
    & {1st } & {2nd} & {3rd}   & {4th}  & {5th}  \\ \midrule
{One-time}       & \ding{51}     & \ding{55}    &-     & -   &    -\\
{Temporary}      &  \ding{51}    & \ding{51}    & \ding{51}      & \ding{55}     &    - \\
{Time-based}     &  \ding{51}    & \ding{51}    & \ding{51}      & \ding{51}    &   \ding{55}         \\
{Invalid Nonce}  &    \ding{51}  &\ding{51}     & \ding{51}      & \ding{51}     &   \ding{51}  \\
\bottomrule
\end{tblr}

\vspace{2pt} 
\raggedright \footnotesize  \ding{51}: Request successful; \ding{55}:Request failed.\\
\end{table}

\paragraph{Signature Checker}
Signature Checker sends three \texttt{QUERY} requests to obtain three different messages. These messages are then used to perform three independent tests: 1) setting the signature to null, 2) setting the signature to an invalid value, and 3) replacing the address with a different one. If the token can be obtained from any of these tests, it indicates that the server does not properly verify the signature or address.

Furthermore, for convenience, an additional \texttt{ACCESS} request is sometimes sent following the \texttt{AUTH} request. 
\checker confirms the vulnerability by checking that the return value of a \texttt{ACCESS} request (instead of \texttt{AUTH} ) is as expected.

\subsection{FlexRequest: A Python HTTP Library}
\label{sec:flexrequest}
To detect vulnerabilities, \checker tests the authentication-related APIs of the website.
Given that the differences in the APIs of each website, existing testing tools or libraries, such as \textit{Postman}\cite{postman} and \textit{Requests}\cite{requests}, require the development of separate test scripts for each site, leading to issues of code duplication and maintainability. 
To address these challenges, we developed a specialized HTTP library for \detector.
\fr, a Python-based HTTP library, features an automatic replacement mechanism to align with the variations in APIs, providing a flexible and adaptable solution for testing APIs on various websites.
As illustrated in Figure~\ref{fig:Web3AuthChecker}, \fr utilizes \textit{cffi\_curl} for HTTP requests. This Python library creates bindings for \textit{curl-impersonate} through the C Foreign Function Interface (CFFI). \textit{curl-impersonate} is a special build of \textit{curl} that can impersonate the four main browsers.
By performing TLS and HTTP handshakes identical to a real browser, \textit{curl-impersonate} ensures that websites do not block requests.

\fr supports using keys to replace dynamic parameters in the API. Before executing a request, it automatically substitutes these keys with the corresponding values. After receiving the response, \fr retrieves the values at the specified positions according to a predefined configuration, binding them to the corresponding keys.
By managing the values of these keys, developers can perform unified testing across various APIs.

Furthermore, \fr maintains a \textit{session context} throughout a session to pass the previous response values between a series of requests. For example, the message obtained from the \texttt{Query} response will be stored in the session context and bound to the key \textit{msg}.
When conducting an \texttt{Auth} request, \fr automatically populates the request with the value of \textit{msg} from the session context.
For detailed insights into API testing challenges and \fr's unique solutions, please refer to our open-source tool, \detector\footnote{\url{https://github.com/d0scoo1/Web3AuthChecker}}.

\begin{table*}[t]
\caption{Detection Results of Websites that Support Web3 Authentication}
\label{tab:results}
\centering
\footnotesize
\SetTblrInner{colsep=2pt,rowsep=0pt,stretch = 1}
\begin{tblr}{
    width = 1\linewidth, 
    colspec ={X[0.5,c]X[2,c]X[1.5,c]X[0.15,c]
    X[0.9,c]X[0.7,c]X[0.7,c]X[0.2,c]
    X[1,c]X[0.7,c]X[0.8,c]X[1.1,c]X[1,c]X[0.2,c]
    X[0.8,c]X[0.8,c]X[1,c]},
    row{odd} = {white},
    row{even} = {gray!25},
    row{1} = {white, font=\bfseries},
    row{2} = {white, font=\bfseries},
    column{1} = {colsep=4pt, font=\bfseries},
    cell{1}{1} = {r=2}{c},
    cell{1}{2} = {r=2}{c},
    cell{1}{3} = {r=2}{c},
    cell{1}{5} = {c=3}{c},
    cell{1}{9} = {c=5}{c},
    cell{1}{15} = {c=3}{c},
}
\toprule
 \#     & Website    & Category   &  &  
Message Design {\color{red}(V1)}  &            &           &           & 
Server Verification  &  &        &      &       & &  
Security Risk    &  \\
        &        &         &           &
Domain  &  Name &  Nonce  & &  
Message & Body & Nonce & Signature & Address  & & 
BMA     & RA & BMMA\\
\midrule
1   & Blur  & Marketplace    & &
{\color{red}\ding{56}}  & {\color{green}\ding{51}} & {\color{green}\ding{51}} & &
{\color{green}\ding{51}} & {\color{green}\ding{51}} & {\color{green}\ding{51}} & {\color{green}\ding{51}} & {\color{green}\ding{51}}& &  
{\bf \color{red}\textsf{M}}	 &  $\Circle$  &  $\Circle$  \\

2   & OpenSea   & Marketplace     & & 
{\color{green}\ding{51}} &    {\color{green}\ding{51}}    & {\color{green}\ding{51}} & &
{\color{green}\ding{51}} & {\color{green}\ding{51}} & {\color{green}\ding{51}} & {\color{green}\ding{51}} & {\color{green}\ding{51}} & &  
{\bf \color{green}\textsf{L}}	 &  $\Circle$  &  $\Circle$  \\

3   & LooksRare   & Marketplace      & & 
{\color{green}\ding{51}} &    {\color{green}\ding{51}}    & {\color{green}\ding{51}} & & 
{\color{green}\ding{51}} & {\color{green}\ding{51}} & {\color{green}\ding{51}} & {\color{green}\ding{51}}& {\color{green}\ding{51}} & & 
{\bf \color{green}\textsf{L}} &   $\Circle$	&  $\Circle$  \\

4   & Foundation     & Marketplace       & &
{\color{red}\ding{56}}  & {\color{green}\ding{51}} & {\color{red}\ding{56}} & &
{\color{green}\ding{51}}  & {\color{red}\ding{56}(V3)}  & \texttt{N/A}  & {\color{green}\ding{51}}  & {\color{green}\ding{51}} & &  
{\bf \color{red}\textsf{M}}   & $\CIRCLE$	& $\CIRCLE$  \\

5   & Element & Marketplace        & & 
{\color{green}\ding{51}} &    {\color{green}\ding{51}}    & {\color{green}\ding{51}} & &
{\color{green}\ding{51}} & {\color{red}\ding{56}(V3)}  & {\color{green}\ding{51}}  & {\color{green}\ding{51}}& {\color{green}\ding{51}}  & & 
{\bf \color{red}\textsf{M}} &	$\Circle$ & $\CIRCLE$  \\

6   & Rarible        & Marketplace    & & 
{\color{green}\ding{51}} &    {\color{green}\ding{51}}    & {\color{green}\ding{51}}& & 
{\color{green}\ding{51}} & {\color{green}\ding{51}} & {\color{green}\ding{51}}& {\color{green}\ding{51}} & {\color{green}\ding{51}}  & &  
{\bf \color{green}\textsf{L}} &  $\Circle$ &  $\Circle$  \\

7   & Joepegs        & Marketplace  & & 
{\color{red}\ding{56}}  & {\color{green}\ding{51}} & {\color{green}\ding{51}} & &
{\color{green}\ding{51}} & {\color{green}\ding{51}} & {\color{green}\ding{51}} & {\color{green}\ding{51}} & {\color{green}\ding{51}} & &  
{\bf \color{red}\textsf{M}} &  $\Circle$&  $\Circle$  \\

8   & Quix  & Marketplace     & & 
{\color{red}\ding{56}}  & {\color{red}\ding{56}}  & {\color{green}\ding{51}} & & 
{\color{green}\ding{51}} & {\color{red}\ding{56}(V2)}  & {\color{red}\ding{56}(V3)}  & {\color{green}\ding{51}}& {\color{green}\ding{51}}  & &  
{\bf \color{red}\textsf{H}}  & $\CIRCLE$	& $\CIRCLE$  \\

9   & Minted Network & Marketplace    & & 
{\color{green}\ding{51}} &   {\color{green}\ding{51}}     & {\color{green}\ding{51}} & &
{\color{green}\ding{51}} & {\color{green}\ding{51}} & {\color{green}\ding{51}} & {\color{green}\ding{51}}  & {\color{green}\ding{51}}  & &  
{\bf \color{green}\textsf{L}} &   $\Circle$	&  $\Circle$  \\

10  & Campfire       & Marketplace  & & 
{\color{green}\ding{51}} &   {\color{green}\ding{51}}     & {\color{green}\ding{51}} & &
{\color{green}\ding{51}} & {\color{green}\ding{51}} & {\color{green}\ding{51}} & {\color{green}\ding{51}} & {\color{green}\ding{51}} &  &  
{\bf \color{green}\textsf{L}} & $\Circle$ &  $\Circle$  \\

11  & Moonflow NFT & Marketplace   & & 
{\color{red}\ding{56}}  & {\color{green}\ding{51}} & {\color{green}\ding{51}} & & 
{\color{green}\ding{51}} & {\color{green}\ding{51}} & {\color{green}\ding{51}} & {\color{green}\ding{51}} & {\color{green}\ding{51}} & & 
{\bf \color{red}\textsf{M}}	 & $\Circle$ &  $\Circle$  \\

12  & Galler& Marketplace      & & 
{\color{green}\ding{51}} &    {\color{green}\ding{51}}    & {\color{green}\ding{51}} & & 
{\color{red}\ding{56}(V2)} & {\color{red}\ding{56}(V2)}  & {\color{red}\ding{56}(V2)} & {\color{green}\ding{51}} & {\color{green}\ding{51}} & &  
{\bf \color{red}\textsf{C}}	 &  $\CIRCLE$	& $\CIRCLE$  \\

13  & PlayDapp     & Marketplace    & & 
{\color{red}\ding{56}}  & {\color{red}\ding{56}}  & {\color{red}\ding{56}}  & & 
{\color{green}\ding{51}} & {\color{green}\ding{51}} & \texttt{N/A}   & {\color{green}\ding{51}}  & {\color{green}\ding{51}} & &  
{\bf \color{red}\textsf{H}}  & $\CIRCLE$&  $\Circle$  \\

14  & Refinable      & Marketplace    & & 
{\color{red}\ding{56}}  & {\color{red}\ding{56}}  & {\color{green}\ding{51}} & &
{\color{green}\ding{51}} & {\color{green}\ding{51}} & {\color{green}\ding{51}} & {\color{green}\ding{51}} & {\color{green}\ding{51}} & &  
{\bf \color{red}\textsf{H}}	 & $\Circle$	&  $\Circle$  \\

15  & Apeiron        & Marketplace    & & 
{\color{red}\ding{56}}  & {\color{red}\ding{56}}  & {\color{green}\ding{51}} & &
{\color{green}\ding{51}} & {\color{green}\ding{51}} & {\color{green}\ding{51}} & {\color{green}\ding{51}} & {\color{green}\ding{51}} & &  
{\bf \color{red}\textsf{H}}	 & $\Circle$&  $\Circle$  \\

16  & Lifty & Marketplace       & & 
{\color{green}\ding{51}} &    {\color{green}\ding{51}}    & {\color{green}\ding{51}} & &
{\color{green}\ding{51}} & {\color{green}\ding{51}} & {\color{green}\ding{51}} & {\color{green}\ding{51}} & {\color{green}\ding{51}} & &
{\bf \color{green}\textsf{L}}	 & 	$\Circle$ &  $\Circle$  \\

17  & LearnBlockchain& Community   & & 
{\color{red}\ding{56}}  & {\color{green}\ding{51}} & {\color{red}\ding{56}} & &
{\color{red}\ding{56}(V2)}  & {\color{red}\ding{56}(V2)}  & \texttt{N/A}    & {\color{green}\ding{51}} & {\color{green}\ding{51}} & &  
{\bf \color{red}\textsf{C}}	 & $\CIRCLE$     & $\CIRCLE$  \\

18  & DappRadar      & Ranking          & & 
{\color{red}\ding{56}}  & {\color{red}\ding{56}}  & {\color{green}\ding{51}} & & 
{\color{green}\ding{51}} & {\color{green}\ding{51}} & {\color{green}\ding{51}}& {\color{green}\ding{51}} & {\color{green}\ding{51}} & &  
{\bf \color{red}\textsf{H}}	 & 	$\Circle$  &  $\Circle$  \\

19  & QuestN& Service         & & 
{\color{red}\ding{56}}  & {\color{green}\ding{51}} & {\color{green}\ding{51}} & &
{\color{green}\ding{51}} & {\color{red}\ding{56}(V2)}  & {\color{green}\ding{51}}  & {\color{green}\ding{51}} & {\color{green}\ding{51}} &  &
{\bf \color{red}\textsf{H}}	 & 	$\Circle$ & $\CIRCLE$  \\

20  & Galxe & Social    & & 
{\color{green}\ding{51}} &    {\color{green}\ding{51}}    & {\color{green}\ding{51}}  & &
{\color{green}\ding{51}} & {\color{green}\ding{51}} & {\color{red}\ding{56}(V2)}  & {\color{green}\ding{51}}& {\color{green}\ding{51}} & &  
{\bf \color{green}\textsf{L}}	 & $\CIRCLE$ &  $\Circle$  \\

21  & Planetix       & Game    & & 
{\color{green}\ding{51}} &    {\color{green}\ding{51}}    & {\color{green}\ding{51}} & &
{\color{green}\ding{51}} & {\color{red}\ding{56}(V2)}  & {\color{red}\ding{56}(V2)}   & {\color{green}\ding{51}} & {\color{green}\ding{51}}&  & 
{\bf \color{red}\textsf{H}}	 & $\CIRCLE$ & $\CIRCLE$  \\

22  & MOBOX        & Game     & & 
{\color{red}\ding{56}}  & {\color{red}\ding{56}}  & {\color{green}\ding{51}} & &
{\color{green}\ding{51}} & {\color{green}\ding{51}} & {\color{green}\ding{51}} & {\color{green}\ding{51}} & {\color{green}\ding{51}} & &  
{\bf \color{red}\textsf{H}}	 &  $\Circle$ &  $\Circle$  \\

23  & Bomb Crypto 2  & Game  & & 
{\color{red}\ding{56}}  & {\color{red}\ding{56}}  & {\color{green}\ding{51}} & &
{\color{green}\ding{51}} & {\color{green}\ding{51}} & {\color{green}\ding{51}} & {\color{green}\ding{51}} & {\color{green}\ding{51}}&  &   
{\bf \color{red}\textsf{H}}	 & 	$\Circle$ &  $\Circle$  \\

24  & Decert& Service        & & 
{\color{red}\ding{56}}  & {\color{green}\ding{51}} & {\color{green}\ding{51}} & & 
{\color{green}\ding{51}} & {\color{green}\ding{51}} & {\color{green}\ding{51}} & {\color{green}\ding{51}} & {\color{green}\ding{51}} & &  
{\bf \color{red}\textsf{M}}	 &	$\Circle$ &  $\Circle$  \\

25  & Paragraph      & Media & & 
{\color{red}\ding{56}}  & {\color{green}\ding{51}} & {\color{green}\ding{51}} & & 
{\color{green}\ding{51}} & {\color{green}\ding{51}} & {\color{green}\ding{51}} & {\color{green}\ding{51}} & {\color{green}\ding{51}} & & 
{\bf \color{red}\textsf{M}}& $\Circle$ &  $\Circle$  \\
\midrule
26 & Campfire       & Marketplace       & & 
{\color{red}\ding{56}}  & {\color{red}\ding{56}}  & {\color{red}\ding{56}}  & &
{\color{green}\ding{51}}  & {\color{green}\ding{51}} & \texttt{N/A}   & {\color{green}\ding{51}} & {\color{green}\ding{51}}  & &  
{\bf \color{red}\textsf{H}}	 & $\CIRCLE$&  $\Circle$  \\

27 & Lifty & Marketplace    & & 
{\color{red}\ding{56}}  & {\color{red}\ding{56}}  & {\color{red}\ding{56}}  & &
{\color{green}\ding{51}} & {\color{green}\ding{51}} & \texttt{N/A}    & {\color{green}\ding{51}} & {\color{green}\ding{51}} & &  
{\bf \color{red}\textsf{H}}	 & $\CIRCLE$&  $\Circle$  \\

28  & NFTmall        & Marketplace    & & 
{\color{red}\ding{56}}  & {\color{red}\ding{56}}  & {\color{red}\ding{56}} & & 
{\color{green}\ding{51}}  & {\color{green}\ding{51}} & \texttt{N/A}   & {\color{green}\ding{51}} & {\color{green}\ding{51}} & &  
{\bf \color{red}\textsf{H}}	 & $\CIRCLE$	&  $\Circle$  \\

29 & Babylons       & Marketplace   & & 
{\color{red}\ding{56}}  & {\color{red}\ding{56}}  & {\color{red}\ding{56}} & & 
{\color{green}\ding{51}} & {\color{green}\ding{51}} & \texttt{N/A}    & {\color{green}\ding{51}}  & {\color{green}\ding{51}} & &  
{\bf \color{red}\textsf{H}}	 & $\CIRCLE$ &  $\Circle$  \\
\bottomrule
\end{tblr}

\vspace{2pt}
\raggedright \footnotesize
In this table, BMA = Blind Message Attack, RA = Replay Attack, BMMA = Blind Multi-Message Attack.\\
In the Web3 authentication process, cases \#1$\sim$\#25 are for user login, while cases \#26$\sim$\#29 are for profile update.\\ 
{\color{green}\ding{51}}: No vulnerability found;~~{\color{red}\ding{56}}: Vulnerability found;~~\texttt{N/A}: Not applicable;~~{\bf \color{red}C}: Critical;~~{\bf \color{red}H}: High;~~{\bf \color{red}M}: Medium;~~{\bf \color{green}L}: Low;~~$\CIRCLE$: Risk exists;~~$\Circle$: No risk.
\end{table*}
%

\section{Findings}
\label{sec:measurement}
This section conducts a comprehensive evaluation of Web3 authentication.
Our analysis uncovers the extensive occurrence of \attacks. Additionally, we delve into two specific cases and evaluate the efficacy of \detector.



\paragraph{Dataset}
Given the limited research on Web3 authentication, no relevant datasets exist.
DappRadar, a well-known Web3 dapp distribution platform, provided a basis for our collection: we selected 18 marketplaces that support Web3 authentication from the top 50 in DappRadar's \textit{Top Decentralized Marketplaces} list~\cite{dappradarRanking}. Additionally, we identified nine websites that support Web3 authentication through Google searches.
Two of these 27 websites required Web3 authentication for user login and profile updates (4 test cases). Among the remaining 25 websites, 23 utilized Web3 authentication solely for user login (23 test cases), and two employed it specifically for profile updates (2 test cases). 
Therefore, we have a total of 29 test cases.
According to DappRadar's statistics, in January 2024 alone, the total transaction volume of 16 websites exceeded 592 million US dollars, and the number of unique active wallets (UAW) surpassed 1.29 million.

Our dataset includes various categories such as games, services, and forums.
Notably, NFT marketplaces are the primary entities of Web3 authentication. 
On the contrary, DEXs (Decentralized Exchanges) and DeFi (Decentralized Finance) typically interact directly with the blockchain, making Web3 authentication unnecessary. 
Our examination of the top 25 DEXs and the top 25 DeFi applications revealed that none employ Web3 authentication; therefore, they have been excluded from our dataset.

\paragraph{Setup}
We thoroughly examined each website and collected the relevant APIs, exporting them as JSON files in Postman format. We also prepared corresponding configuration files for \detector.
We deployed \detector on Github Codespaces, configured with 4 cores and 8GB RAM. 
We set Chrome (v110) as the \textit{curl-impersonate} browser and limited the API response timeout to 10 seconds. 
To avoid being blocked due to frequent requests, we instituted a one-minute interval between each request. 


\subsection{Analysis of Results}
Table~\ref{tab:results} provides detailed information and detection results for all 29 test cases.
\detector examines whether messages include \fdomain, \fname, and \fnonce, and it verifies if the server checks the message, message body (body), nonce, signature, and address to identify the risks defined in Section~\ref{sec:attacks}. 
The results are alarming, with 22 out of the 29 test cases found to be subject to medium to critical risks, making them vulnerable to \attacks.
At the same time, we identified critical risks in two cases, and we have informed the respective vendors. One of them has already acknowledged the risk and fixed it.
Also, 11 cases are at the risk of \rattacks, and seven have \mattacks.
Table~\ref{tab:results} shows that each website has its own distinct vulnerabilities, indicating that their Web3 authentication implementations are independent. 
Therefore, we conclude that these websites have no common server-side SDKs.


Subsequently, we will analyze the security risks of 25 cases of using Web3 authentication for user login (\#1$\sim$\#25). 
Then, we will address the four remaining cases (\#26$\sim$\#29) that employ this authentication method for profile updates (note that two websites are included in both categories).


\paragraph{Blind Message Attack}
In the 25 user login cases (\#1$\sim$\#25), the \textit{BMA} column of Table~\ref{tab:results} reveals that two of them are at critical risk ({\bf \color{red}C}), nine at high risk ({\bf \color{red}H}), seven at medium risk ({\bf \color{red}M}), and only seven are evaluated as low risk ({\bf \color{green}L}). 
While all cases verify the signature and address, two cases do not perform any checks on the message, thus resulting in critical risks. 
In two critical-risk cases, an attacker could use a user's arbitrary signature to log in to the user's account.
One of these cases, learnblockchain, has over 2 million users, highlighting a significant risk.

In nine high-risk cases, either the \fdomain and \fname fields do not exist in the message (V1), or the server does not verify the message body so the attacker can tamper with them (V2).
The nine high-risk websites provide attractive targets for attackers. 
By exploiting the users' inability to identify the source of messages, malicious websites can effortlessly trick users into signing messages from these websites.
Notably, Apeiron (\#15), QuestN (\#19), and MOBOX (\#22), all identified as high-risk websites, reported 89.21k, 54.94k, and 27.26k Unique Active Wallets (UAW), respectively, in January 2024.
These UAWs underscore the substantial user engagement and the potential impact of such vulnerabilities.

In the seven medium-risk cases, the attacker may need more sophisticated techniques to implement attacks, such as using a domain name identical to the target website (V1) or hiding the target website's message body within the malicious website's message (V3). 
As discussed in the case study section (Section \ref{sec:case study}), these hurdles may be relatively easy for experienced attackers.
Importantly, even cases categorized as low risk are not necessarily exempt from \attacks, as attackers still have opportunities to deceive unwary users.

Vulnerabilities in verification can weaken a well-structured design.
For instance, Element (\#5), Galler (\#12), and Planetix (\#21) all have appropriately structured messages containing the essential fields. However, Element does not improperly verify the message body (V3), thus placing it at medium risk. 
In contrast, the other two do not verify the message body (V2), which puts them at high risk. 
As of January 2024, Element reported 158.63k UAWs and a transaction volume of \$18.91M. These security risks emphasize the importance of server verification.

\begin{table}[t]
\caption{Nonce Types of the 25 Websites}
\label{tab:nonces}
\centering
\footnotesize
\SetTblrInner{colsep=6pt,rowsep=0pt,stretch = 1}
\begin{tblr}{
    width = 1\linewidth, 
    colspec ={X[2.5,l]X[1.5,c]X[5,c]X[1,c]},
    row{even} = {white},
    row{odd} = {gray!25},
    row{1} = { white, font=\bfseries},
    column{1} = {font=\bfseries},
}
\toprule
{Nonce Type}       & {Replay} & {Websites} & {Total} \\ \midrule
{One-time}         & $\Circle$ &     \#2, \#3, \#5, \#14, \#16, \#24     & 6   \\
{Temporary}        & $\LEFTcircle$      		&  \#1, \#7, \#10, \#11, \#18, \#25        & 6      \\
{Time-based}       & $\LEFTcircle$      		&  \#6,  \#9, \#15, \#19, \#22, \#23  & 6     \\
{Invalid Nonce}      & $\CIRCLE$    &  \#8, \#12, \#20, \#21  & 4   \\
{No Nonce}         & $\CIRCLE$    &  \#4, \#13, \#17        & 3     \\ 
\bottomrule
\end{tblr}

\vspace{2pt}
\raggedright \footnotesize
In this table, $\CIRCLE$:Risk exist; $\LEFTcircle$: Short-term risk; $\Circle$: No risk.
\end{table}

\paragraph{Replay Attack}
In the 25 user login cases (\#1$\sim$\#25), as indicated in the \textit{RA} column of Table~\ref{tab:results}, seven out of these 25 cases are vulnerable to replay attacks. This vulnerability implies that an attacker could reuse a signature, thereby bypassing the session's security mechanism. 
Three cases can be attributed to the message not containing \fnonce (V1). 
The remaining four cases do include \fnonce, but it is either unverified on the server ({\color{red}\ding{56}(V2)}) or the server check contains a vulnerability ({\color{red} \ding{56}(V3)}), as shown in the \textit{Nonce} column of Table~\ref{tab:results}. For instance, Quix (\#8) uses a timestamp as a nonce. If the timestamp in the message is ahead of the current time, Quix's server rejects it. However, the servers do not define an expiration time, meaning used messages and signatures remain valid.

Further investigation into the nonces used by these cases is presented in Table~\ref{tab:nonces}. Among them, 12 cases employ temporary and time-based nonces. These nonces could technically be replayed within their valid period; however, we measured their expiration time and found that they typically have short validity periods, often not exceeding fifteen minutes. As a result, we conclude that these cases are not at significant risk of replay attacks.

\paragraph{Blind Multi-Message Attack}
There are five cases failed to verify the message body ({\color{red}\ding{56}(V2)}), while two cases improperly checked the message body ({\color{red}\ding{56}(V3)}), as shown in the \textit{Body} column of Table~\ref{tab:results}. 
For the former, the high-risk scenario allows an attacker to construct a message body freely, removing crucial details such as the \fdomain and \fname. 
In the latter scenario, characterized as medium-risk, the message body is improperly checked, leaving an opportunity for the attacker to exploit this vulnerability by adding misleading information or cleverly hiding the original message. 
Overall, these two vulnerabilities could be exploited for blind multi-message attacks, suggesting that all seven websites are at such risk as shown in the \textit{BMMA} column of Table~\ref{tab:results}.

\paragraph{Profile Update}
In the four profile update cases (\#26$\sim$\#29), all four cases are vulnerable to high-risk blind message attacks and replay attacks.
NFTmall (\#28) and BabyIons (\#29) do not require user login, but they require Web3 authentication when users update their profiles.
Campfire (\#26) and Lifty (\#27) require additional Web3 authentication for profile updates even after the user has logged in. 
However, Campfire and Lifty do not verify the tokens obtained from the prior login during profile updates. As a result, the security of profile updates relies entirely on the current Web3 authentication implementation.
Both Campfire and Lifty employ well-structured messages without any evident security vulnerabilities during user login. Regrettably, they fail to maintain this level of security in the profile update. 
For instance, Campfire uses the simple phrase \textit{update\_profile} as the message for profile update.
If an attacker obtains a user's signature for this phrase, they could freely update the user's profile.

\subsection{Case Study}
\label{sec:case study}
This section presents two case studies highlighting the security issues. 
The first case study examines the unchecked message vulnerability (V1) 
and has been acknowledged by LearnBlockchain.
The second case study underscores a blind multi-message attack, demonstrating how it is possible to bypass Web3 authentication on three distinct websites utilizing just a single message.

\paragraph{Unchecked Message}
LearnBlockchain (\#17) is a well-known blockchain community with more than 2 million users. It supports both traditional password-based authentication and Web3 authentication. In the Web3 authentication, LearnBlockchain's message is simply \textit{learnblockchain}. Moreover, the server only verifies the signature and does not perform any checks on the message, allowing any signature of the user to pass the Web3 authentication. This vulnerability provides an easy way for an attacker to gain unauthorized access to a user account, which can lead to the exposure of important personal information and potential asset losses. 
Specifically, LearnBlockchain’s points system supports points exchange for currency, so attackers can transfer points from the victim’s account to their own account and then withdraw them. 

We promptly reported blind message attack and replay attack to LearnBlockchain, who fixed their Web3 authentication based on our suggestions and awarded us 2500 points
for our responsible disclosure.
Another website, Galler (\#12), facing similar security issues, was also notified about the vulnerabilities. However, at the time of writing, we have not received any response from Galler.

\paragraph{Blind Multi-Message Attacks}
\label{sec:blind multi-message attacks}
The vulnerabilities associated with message body verification can cause more extensive damage than expected. 
An attacker can create a message while bypassing the Web3 authentication of multiple websites.
In this way, attackers can steal a user's corresponding identity on multiple websites at once. 
We refer to this situation as a {\textsf{\small Blind Multi-Message Attack}} (BMMA).

\begin{lstlisting}[language=ABNF, caption={A Malicious Message For the BMMA}, label={lst:malicious message}]
/** Modified based on Foundation's message. **/
Welcome! Please sign this message to connect to Foundation.com.

/** Planetix **/
Web3 Token Version: 2 
Nonce: 84800972
Issued At: 2024-01-13T03:59:00.000Z
Expiration Time: 2024-01-14T03:59:00.000Z

/** QuestN **/
Timestamp: 1706389762 
\end{lstlisting}

Listing~\ref{lst:malicious message} presents an example of a malicious message that can bypass the authentication of three different websites: Foundation (\#4), QuestN (\#19), and Planetix (\#21). 
The attacker first registers a domain on \textit{foundation.com}. 
Exploiting the vulnerability of \textit{foundation.app}, which only checks for the existence of a message body without verifying its content, the attacker cleverly appends \textit{Welcome!} to the beginning of the foundation.app's message (Listing~\ref{lst:bad message}) and \textit{com.} to the end. 
Additionally, Planetix and QuestN do not check the message body. Planetix verifies four fields: \textit{Web3 Token Version}, \textit{Nonce}, \textit{Issued At}, and \textit{Expiration Time}, while QuestN merely verifies the timestamp as \fnonce. The attacker appends fields required by Planetix and QuestN to the end of the message.
The malicious message is obviously different from the messages on the three websites, and it contains the \fdomain of foundation.com, thereby misleading users into believing it originates from the malicious website (foundation.com).
As a result, users trust that the message is safe and easily sign the message on the malicious website.
This example underscores the severity of blind multi-message attacks.

Besides, the Foundation's message lacks a nonce, implying that once the attacker acquires a user's signature, they can maintain unauthorized access through replay attacks.

\subsection{Evaluation of \detector}
\detector tests each attack payload from scratch every time, with multiple payloads per vulnerability. Therefore, in the experiment, \detector sent a total of 1,319 requests, an average of 44 requests per website.
We manually checked all the results of \detector and found that it did not produce any false positives or false negatives. We also compared \fr's performance with that of the widely-used Python HTTP library, \textit{Requests}~\cite{requests}, and the tool \textit{Postman}~\cite{postman}. The results indicated that six websites rejected requests from both \textit{Requests} and \textit{Postman}, even when the request headers were modified to mimic other browsers. In contrast, \fr successfully completed requests to all APIs of these websites, demonstrating its superior capability in handling APIs from a diverse range of websites.


\section{Mitigation: Web3AuthGuard}
\label{sec:mitigation}
Given the vulnerabilities identified by \detector, extensive server-side updates are necessary but challenging to implement quickly. This section introduces \guard, a solution designed to detect potential attacks on the user (wallet) side.
Compared to \detector, \guard automatically detects potential attacks during wallet operation (signature request) without the need for pre-configuration.

\subsection{Design of \guard}

\begin{figure}[t]
    \centering
	\includegraphics[width=0.95\linewidth]{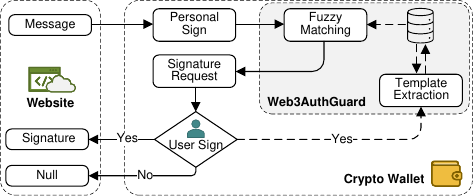}  
	\caption{Workflow of Web3AuthGuard.}
	\label{fig:Web3AuthGuard}
\end{figure}

\guard is designed to detect suspicious messages in crypto wallets, to prevent \attacks during the Web3 authentication process. The workflow of \guard is depicted in Figure~\ref{fig:Web3AuthGuard}.
During the Web3 authentication process, a website triggers the wallet's signature API (\textit{Personal Sign}). Consequently, the wallet prompts a \textit{Signature Request} page for the user to sign the message.
For wallets featuring \guard, the first step involves comparing the new message with previously signed messages (message templates). If the new message significantly resembles a previously signed message from a different website, \guard promptly raises an alert on the top of the signature request page. 
This alert notifies the user of a potential \attack, enabling them to decide whether to proceed with the authentication process. 
The above process mainly relies on three components of \guard: \textit{template extraction}, \textit{fuzzy matching}, and \textit{wallet alerts}. Below, we introduce them in detail.

\paragraph{Template Extraction}
Each time the user logs in, \guard extracts the message template from the message. It then stores this template in the wallet with the domain name for subsequent fuzzy matching.
We avoid directly storing the message, as it often contains variable fields (such as \fnonce). \guard extracts a message template from multiple messages originating from each website. 
The template captures the static fields of the message, substituting variable fields with wildcards.

Template extraction process is as follows: when the user logs into the website for the first time, the wallet will record the message and mark it with the domain name. When the user logs in to the website again, \guard will compare the new message with the previously stored message to extract the message template. Specifically, the process involves a word-by-word comparison of the two messages. Words that match are kept and incorporated into the template. In contrast, non-matching words, pinpointed by predefined variable fields, are substituted with corresponding wildcards in the template. This updated template then supersedes the previously saved message and becomes the reference for detecting blind message attacks.

The template extraction is a dynamic and adaptive process; it can capture changes when a website's message is updated, eliminating the need for manual intervention. \guard only retains the latest template for each website. A few megabytes are generally sufficient for wallets. In experiments, the total size of templates for 25 websites was less than 10 KB, with the largest being 1.21 KB. Consequently, 1MB can accommodate at least 800 such templates.

\begin{figure}[t]
\label{fig:alert}
\centering  	
\subfigure[Blind Message Attack]	
{\label{fig:alert1}
    \begin{minipage}[t]{0.47\linewidth}
        \centering        
        \includegraphics[width=1\linewidth]{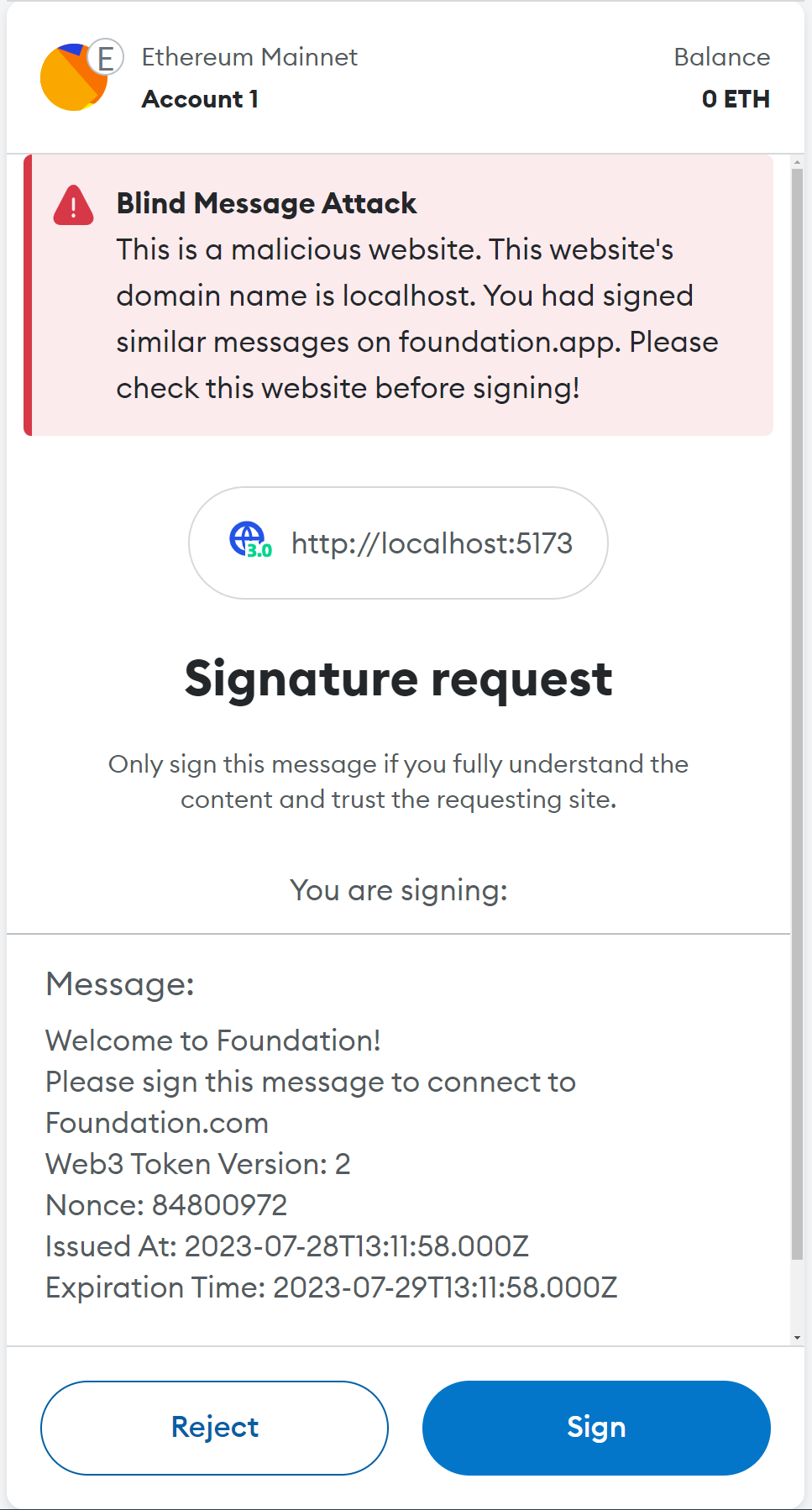}  
    \end{minipage}
}
\subfigure[Lack of Domain Name]
{
    \label{fig:alert2}
    \begin{minipage}[t]{0.47\linewidth}
        \centering     
        \includegraphics[width=1\linewidth]{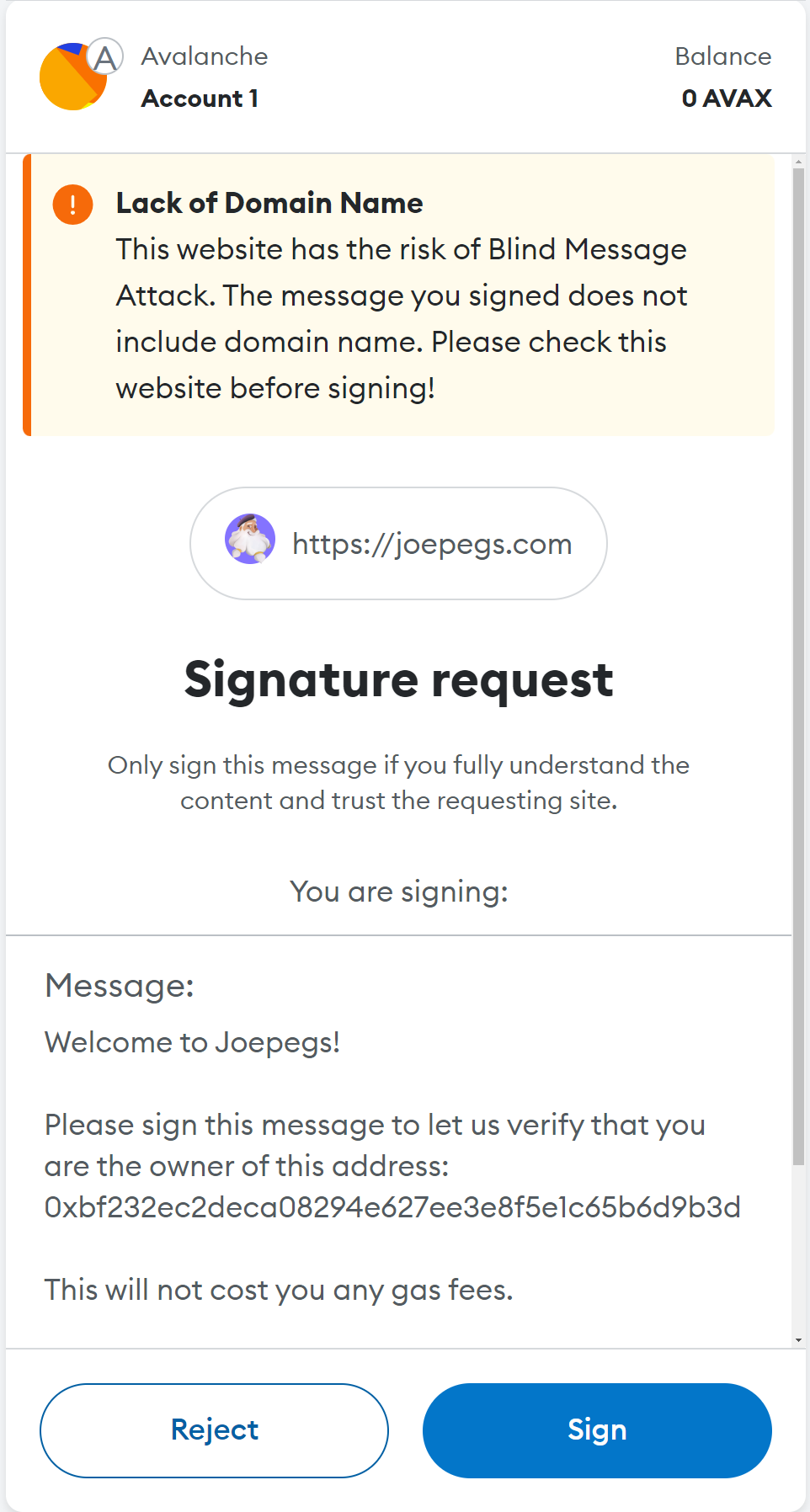}
    \end{minipage}
}
\caption{Alerts in Signature Requests.}
\label{fig:alerts}
\end{figure}

\paragraph{Fuzzy Matching}
\guard employs regular expression matching to compare the new message with the stored message template during the authentication process. If a match is found, indicating similarity between the new message and the template, \guard further checks if they originate from different websites. If they do, \guard adds an alert on the signature request page, indicating a potential blind message attack. 

\paragraph{Wallet Alerts}
\guard will add alerts to the signature request page to inform users of potential security risks.
There are two types of alerts, \textit{blind message attack} and \textit{lack of domain name}, which are independent and may occur at the same time.
If fuzzy matching detects a blind message attack,
\guard will add a red alert,
as shown in Figure~\ref{fig:alert1}.
The alert will prompt for the domain names of the current and victim websites.
The user should check whether the domain name (\fdomain) or the similar website name (\fname) of the victim website exists in the message.
If it exists, it is a blind message attack, and the user should click the \textit{Reject} button on the signature page to refuse the signature.
If the \fdomain and \fname do not exist in the message, this may be a potential attack.
Once the user signs, the website will \textit{intentionally} or \textit{unintentionally} gain access to the user's account on the victim's website.
Therefore, we still recommend that users do not sign it. 

In addition, \guard will match whether the \fdomain of the visited website appears in the message. 
If it does not exist, 
\guard will raise a yellow alert (\textit{lack of domain name}), as shown in the figure ~\ref{fig:alert2}.
We recommend that users do not sign the message as it may originate from other websites or be exploited by other malicious websites.
Finally, users should report any malicious activity on the website to their wallets, and most wallets will then warn other users about these dangerous websites.

\subsection{Evaluation of \guard}
To evaluate the effectiveness, we implemented \guard in the open-source code of the MetaMask browser extension for Chrome, the most widely used crypto wallet with over 30 million monthly users~\cite{metamask-stat}.
Our evaluation involved 25 websites (\#1$\sim$\#25) listed in Table \ref{tab:results} that employ Web3 authentication for the user login. We excluded four cases of updated profiles because the content of the messages in these cases was the updated profile rather than the standard message fields.

For the 25 websites, we collected two sets of messages: one for template extraction (comprising five messages) and another for testing (also comprising five messages). Initially, we embedded the templates extracted after four rounds into MetaMask and started testing.
We evaluated \guard's effectiveness by observing if it generated an alert on the MetaMask signature page during Web3 authentication on a malicious website.

To ensure a comprehensive evaluation, we set up a local malicious website capable of sending signature requests to MetaMask and conducted three rounds of testing:
In the first round, the malicious website directly used a second set of five messages from each website to send signature requests, testing whether \guard could detect blind message attacks.
In the second round, 
we assessed whether \guard could identify attacks in the presence of verification flaws;
the malicious website randomly embedded each test message into a message from another website before sending signature requests.
In the third round, we directly visited 25 websites and performed Web3 authentication to evaluate whether \guard generated false positives.

In the 25 user-login cases, \guard successfully passed all three rounds of testing without false positives or negatives. These tests encompass low-risk, medium-risk, and certain high-risk scenarios, thoroughly examining \guard's effectiveness across various potential threats.

\paragraph{Limitation}
Our evaluation shows that \guard is robust without false positives or negatives.
In addition, \guard effectively handles server-side {\it verification flaws} (V3), and its fuzzy matching can detect attacks even when malicious websites embed others' messages into their own.
However, \guard is not entirely free of false positives or negatives.
In two situations, \guard might produce false positives (\attack) in cases where messages from two different websites, lacking \fdomain and \fname fields, have the same message structures.
Also, \guard is ineffective against the vulnerability {\it unchecked fields} (V2); this vulnerability allows attackers to alter the message content (message body) at will, rendering any user-side detection ineffective.
In our testing of 25 cases, five exhibited this vulnerability, as detailed in the
{\it body} column of Table~\ref{tab:results}. 
Consequently, we have manually classified these five cases as failures.
In conclusion, \guard was effective in detecting \attacks on 20 of the 25 tested websites.
Another limitation is that \guard requires multiple messages from the same website to extract a valid template when the website's message contains variable fields. A practical solution is to preload templates of popular websites into crypto wallets.

\section{Discussion}
\label{sec:discussion}
\paragraph{Comparing Threats in Web2}
Ensuring the security of the Web3 ecosystem is paramount for maintaining trust in the decentralized vision, and \attacks represent a significant threat.
Similar to traditional Web2 authentication threats like phishing attacks and man-in-the-middle (MitM) attacks, the ultimate goal of blind message attacks is to gain unauthorized access by deceiving users.
However, unlike phishing attacks, which masquerade as the target website, or MitM attacks that intercept or alter communications, blind message attacks exploit vulnerabilities in the Web3 authentication process. This exploitation offers greater flexibility and effectiveness in deceiving users, as a single malicious website can target different users with varying websites.

\paragraph{Protocol Flaws \& Solutions}
The main reason for the flaws in the Web3 authentication protocol is that the website uses EIP-191~\cite{ERC-191} for signature. The EIP-191 protocol is not specifically designed for Web3 authentication but is only used to implement the \textit{signature} function.
EIP-191 allows a user to sign arbitrary content as the message for Web3 authentication, and the client (user or wallet) cannot effectively confirm whether the message was issued by the visiting website, which allows malicious websites to launch \attacks.
We propose two solutions to address the flaws in the Web3 authentication protocol, including 1) designing a new end-to-end protocol or 2) using unique public keys on each website.
These two solutions are orthogonal.

\sparagraph{1) New Protocol}
The EIP-191 protocol was not designed for Web3 authentication, so a new security protocol is needed. This new protocol should cover both the client and server and meet the following security policies:
\begin{itemize}[leftmargin = 12pt]
    \item \textit{Verifiable Source.}
    The client can easily verify the source of the message. Meanwhile, each website's message should be unique and cannot be impersonated.
    \item \textit{Integrity.} The server can easily verify the integrity of a message to ensure that it was issued by itself and has not been modified.
\end{itemize}
To implement the above security policy,
the new protocol should mandate the presence of \fdomain and \fnonce security fields in the message. At the same time, the protocol should specify the message format so that the client and server can automatically parse these fields.
During the Web3 authentication process, the client (wallet) can automatically check these security fields in the message and prompt the user of potential security risks.
During the verification process, except to verify the validity of the signature, the server also parses the fields in the message. It verifies them one by one to ensure the integrity of the message.
Although the end-to-end protocol can effectively solve the flaws in the current Web3 authentication protocol, it is difficult to implement in the short term because it requires upgrading both the server and the wallet.

\sparagraph{2) Unique Identity}
The private key can generate multiple public keys, allowing users to use a unique key for Web3 authentication on each website, effectively eliminating blind message attacks. 
This also enhances user anonymity, as attackers cannot link a user's public key to all their visited websites.
This solution can be implemented by upgrading the wallet. When a user uses Web3 authentication, the wallet checks for an existing public key for the website and generates a new one if none exists.
However, this approach isolates the user's digital assets across different websites. This issue can be addressed with account abstraction (EIP-4337)~\cite{eip-4337}, although it requires additional transaction fees.

\paragraph{Ethical Concerns}
Our study conducted experiments in the wild, which may raise ethical concerns. 
We critically analyzed our work using the ethics framework~\cite{DBLP:conf/uss/KohnoAL23} to assess our study's ethical considerations and potential risks.
During the experiment, we used randomly generated accounts from a wallet to detect vulnerabilities in Web3 authentication; no normal users were affected. 
The APIs and functions we tested are all public functions of the websites, and we actively notified the websites with vulnerabilities and offered suggestions for improvements. 
To demonstrate the attacks, we built non-public websites and used randomly generated accounts representing both attackers and victims. 
We conducted the demonstration locally to ensure that no malicious websites were created online to mislead or deceive users.
We adhered to Responsible Disclosure, and because some websites had not yet responded, we submitted our security report to the CVE database. We understand that once vulnerabilities in decentralized services are disclosed, attackers can exploit them. Therefore,
we will only release the code of \guard and not disclose the test scripts of websites.

\section{Related Work}
\label{sec:related work}
To the best of our knowledge, we are the first to perform an in-depth study of security in Web3 authentication.
Our research relates to Web authentication and distributed digital identity.

\paragraph{Web2 User Authentication}
Web2 Authentication security focuses on verifying user identities to prevent unauthorized access, which includes Password-based Authentication~\cite{DBLP:conf/uss/MunyendoMNGKUA22, DBLP:conf/sp/PalD0R19, DBLP:conf/sp/MaYLL14}, Multi-Factor Authentication (MFA)~\cite{DBLP:conf/uss/GollaHLPR21, DBLP:conf/sp/ReynoldsSRDRS18, DBLP:conf/sp/KrombholzBP0Z19}, 
Single Sign-On (SSO)~\cite{DBLP:conf/ccs/JannettMMS22, DBLP:conf/uss/GhasemisharifRC18, DBLP:conf/uss/CalzavaraFMSST18}, 
and Fast IDentity Online 2 (FIDO2)
~\cite{DBLP:conf/uss/LassakHGU21, DBLP:conf/sp/HanzlikLW23, DBLP:conf/sp/LyastaniSN0B20}.
Despite its security criticisms, password-based authentication remains prevalent due to its simplicity.
Multi-Factor Authentication, particularly two-step verification, increases password-based authentication's security~\cite{DBLP:conf/sp/KrombholzBP0Z19} by requiring supplemental factors like a one-time password via SMS or email.
Single Sign-On streamlines authentication by enabling a single credential's use across numerous platforms. 
Common SSO standards include OpenID Connect (OIDC)~\cite{DBLP:conf/eurosp/MainkaMSW17}, 
OAuth 2.0~\cite{DBLP:conf/raid/PhilippaertsPJ22}, 
and SAML~\cite{DBLP:conf/uss/SomorovskyMSKJ12}. 
FIDO2, which uses hardware tokens, offers an attractive alternative to password-based authentication. 
Lyastani et al.~\cite{DBLP:conf/sp/LyastaniSN0B20} provide an overview of user authentication technologies, focusing on the potential of FIDO2 for passwordless authentication.

Web2 authentication is vulnerable to common security threats like password guessing~\cite{DBLP:conf/sp/PasquiniGABC21, DBLP:conf/ccs/XuWYZZH21, DBLP:conf/uss/PasquiniCAB21, DBLP:conf/eurosp/YuM22}, 
session hijacking~\cite{DBLP:conf/ccs/DrakonakisIP20, DBLP:conf/sp/GhasemisharifKP22, DBLP:conf/uss/SudhodananP22}, phishing~\cite{DBLP:conf/sp/LainKC22}\cite{DBLP:journals/compsec/AlhogailA21}, 
man-in-the-middle (MitM) attacks~\cite{DBLP:conf/uss/Birge-LeeWMSRM21}\cite{DBLP:conf/uss/KarapanosC14}, 
and credential stuffing~\cite{DBLP:conf/uss/PalIBSVWWR022}\cite{DBLP:conf/uss/WangR21}. 
Web3 authentication, which is based on cryptography and effectively reduces threats such as password guessing, may still be vulnerable to session hijacking since it commonly uses traditional session management.
Blind message attacks pose a unique challenge to Web3, as they can automatically select target websites based on address and blockchain data, unlike phishing, which can only mimic a specific website. Moreover, they do not require interception of communication (MitM), but rather bypass authentication by stealing user identities.


\paragraph{Distributed Digital Identity}
Decentralized Identifiers (DIDs) are user-generated identifiers that circumvent the need for a centralized registry
~\cite{DBLP:journals/jnca/LiuHOKKC20}\cite{DBLP:journals/fgcs/LiaoGCY22}.
DIDs play a crucial role in the Web3 ecosystem, where through public key authentication technology, Web3 applications can easily verify user identities. 
Other techniques, such as zero-knowledge proofs~\cite{DBLP:journals/joc/FeigeFS88}\cite{DBLP:journals/compsec/YangL20} also hold relevance in this field.
MPCAuth~\cite{DBLP:conf/sp/TanCDP23} is a multi-factor authentication system for distributed-trust applications that addresses ease-of-use and privacy challenges.
Ansaroudi et al.\cite{DBLP:conf/dbsec/AnsaroudiCSR23} systematically analyze multiple digital identity wallets.
Korir et al.\cite{DBLP:conf/soups/KorirPD22} conducted a study on decentralized identity wallets, revealing user misconceptions about DIDs.




\paragraph{Web3 Authentication Protocol}
Sign-In with Ethereum (SIWE, EIP-4361)\cite{ERC-4361} requires the inclusion of specific fields such as \fdomain, \fnonce in the message, and crypto wallets supporting the SIWE protocol will verify these fields. Our research complements SIWE; of the 27 websites we examined that use the SIWE protocol (\#10, \#20), one (\#20) had a replay attack risk.
Some EIPs (Ethereum Improvement Proposals) provide signed data and identity standards. EIP-191~\cite{ERC-191} and EIP-712~\cite{EIP-712} improve message readability and complex data signing, respectively, while EIP-725~\cite{EIP-725} suggests a multi-key control scheme for proxy smart contracts. However, they have yet to gain widespread adoption due to issues like distributed server incompatibility (SIWE) and transaction fees (EIP-725).
\section{Conclusion}
\label{sec:conclusion}
This paper investigates the security risks associated with Web3 authentication and introduces the concept of \attack. This attack enables attackers to gain unauthorized access to user accounts. Meanwhile, we propose advanced attacks, namely \rattack and \mattack. Our dynamic detection tool, \detector, successfully identifies 22 out of 29 real-world deployments of Web3 authentication that are at risk of \attacks. 
To better alert users,
we have developed \guard, a client-side solution to detect \attacks in crypto wallets,
and showed that it can alert users immediately of potential attacks.
\detector and \guard are open source to facilitate future research.
\begin{acks}
We thank the anonymous reviewers and our anonymous shepherd for their comments and suggestions.
The authors from Shandong University were supported in part by Taishan Young Scholar Program
of Shandong Province, China (Grant No. tsqn202211001) and Xiaomi Young Talents Program.
This research was also supported by an Ethereum Foundation Academic Grant.
\end{acks}

\balance


\bibliographystyle{ACM-Reference-Format}
\bibliography{refs-intro,refs-related,refs-misc}

\appendix

\section{Extension Field of the Message}
\label{app:extension field}

The extension fields, as shown in Listing~\ref{lst:extension fields}, include many useful optional fields with limited security impact on Web3 authentication. Here, we briefly explain them.

\begin{lstlisting}[language=ABNF, caption={Extension Fields}, label={lst:extension fields}]
address = "0x" 40*40HEXDIG
  ; Must also conform to captilization
  ; checksum encoding specified in EIP-55 (EOAs).

version = "1"

chain-id = 1*DIGIT
  ; See EIP-155 for valid CHAIN_IDs.

issued-at = date-time

expiration-time = date-time

not-before = date-time

request-id = *pchar
  ; See RFC 3986 for the definition of "pchar".
\end{lstlisting}

\paragraph{address}
Listing~\ref{lst:good message} has an address field to tell the user which account to sign the message. However, the address field is unnecessary because the crypto wallet UI has more straightforward prompts, usually including account avatars and aliases. 

\paragraph{version} The version field indicates the version of the message format. Also, the server can determine the version of the message by parsing the message format.

\paragraph{chain-id}
The chain-id field indicates the chain ID of the blockchain network. The server can determine the chain ID of the message by parsing the message format. The chain ID is usually used to prevent replay attacks across different blockchain networks. For example, if the user signs a message on the Ethereum \textit{testnet}, the attacker cannot use the signature to invade the user's account on the Ethereum \textit{mainnet}.

\paragraph{issued-at, expiration-time, not-before-time}
The three fields prompt the validity period of the message.
The issued-at field is a datetime string of the current time.
The expiration-time field indicates when the signed message is invalid.
The not-before-time field indicates when the signed message becomes valid. 
The server must have the corresponding verification if these fields appear in a message.

\paragraph{request-id}
Users may have multiple devices to log in to the website using Web3 authentication. The request-id field is a system-specific identifier that uniquely refers to the login request.

\section{Further Details on FlexRequest}
\label{app:flexRequest}
To illustrate the challenges of API testing, consider the \texttt{QUERY} responses and \texttt{AUTH} requests from two different websites as shown in Listing~\ref{lst:request example1} and Listing~\ref{lst:request example2}. In the case of galler.io, the \texttt{QUERY} returns a message directly, whereas, for element.market, the \texttt{QUERY} only returns a nonce. The \texttt{AUTH} request parameters also differ in both examples, requiring distinct test scripts for each website. This would typically lead to considerable code duplication and challenges in maintaining the codebase.

\begin{lstlisting}[language=json, ,firstnumber=1, caption={The Response and Request of galler.io}, label={lst:request example1}]
QUERY Response:
{'data':{'auth':{'message':'This is Galler, welcome...
timestamp: 1625468800000'}}}

AUTH Request:
{method:'POST', url:'https://www.galler.io/api/v1',
headers:{...}, data:{address:'{{ addr }}',
message:'{{ msg }}', signature:'{{ sig }}'}}
\end{lstlisting}

\begin{lstlisting}[language=json, ,firstnumber=1, caption={The Response and Request of element.market}, label={lst:request example2}]
QUERY Response:
{'data':{'auth':{'nonce':'3deca92b'}}}

AUTH Request:
{method:'POST', url:'https://api.element.market/graphql',
headers:{'x-viewer-addr':'{{ addr }}',...}, data:{message:
'{{ msg }}', nonce:'{{ nonce }}', signature:'{{ sig }}'}}
\end{lstlisting}

\begin{figure}[t]
\centering
\includegraphics[width=0.9\linewidth]{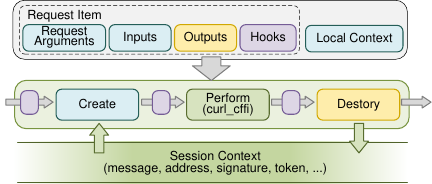}
\caption{FlexRequest in Operation.}
\label{fig:flexRequest}
\end{figure}

However, FlexRequest's automatic replacement mechanism harmonizes these differences. As seen in Listings~\ref{lst:request example1} and~\ref{lst:request example2}, keys such as \textit{addr}, \textit{msg}, etc., are set in the \texttt{Auth} requests. By managing the values of these keys, developers can perform unified testing across multiple APIs.
For instance, to test whether a token can still be obtained with an incorrect signature, one needs to set the value of the \textit{sig} key to be empty, which would put an empty signature in all \texttt{Auth} requests.

\paragraph{FlexRequest in Operation}
As shown in Figure~\ref{fig:flexRequest}, FlexRequest operates in three phases for each request:

\begin{itemize}[leftmargin = 12pt]
\item \textsc{Create.} FlexRequest substitutes keys in the request with values from the \textit{local context}, \textit{session context}, and \textit{inputs}.
\item \textsc{Execute.} FlexRequest carries out the request using \textit{curl\_cffi}.
\item \textsc{Destory.} FlexRequest retrieves values from the specified position in the response according to \textit{outputs}, and stores them in \textit{the session context} in a key-value format.
\end{itemize}

A Request Item encapsulates all the pertinent details of a request, including the URL, headers, and so forth. It also includes two special parameters, \textit{inputs} and \textit{outputs}.
Developers can define default values (\textit{inputs}) and return values (\textit{outputs}) in the API's configuration file and set test values in each checker's \textit{local context}. The local context is only valid for the current request, and those values are first filled into the request.

The key-value replacement function of FlexRequest is very powerful, and the value of a key can even be an executable expression. FlexRequest uses Python's \textit{eval} function to evaluate expressions and return results. Therefore, developers can even simulate front-end JavaScript execution by executable expressions.
Besides, FlexRequest also provides hooks for complex API testing.

\end{document}